\documentclass[12pt]{article}
\pdfoutput=1
\usepackage[utf8]{inputenc}
\usepackage[T1]{fontenc}
\usepackage{float}
\usepackage{authblk}
\usepackage[normalem]{ulem}
\usepackage{xcolor}
\usepackage{amsmath}
\usepackage{amsthm}
\usepackage{amssymb}
\usepackage{amsfonts}
\usepackage{adjustbox}
\usepackage{bbm}
\usepackage{graphicx}
\usepackage{booktabs}
\usepackage{mathtools}
\usepackage{multirow,array}
\usepackage{fullpage}
\usepackage{comment}
\usepackage[mathlines]{lineno}
\usepackage{multirow}
\usepackage{subcaption}
\usepackage{bm}
\usepackage{parskip}
\usepackage[style=numeric-comp,sorting=none,maxnames=99]{biblatex}

\usepackage[colorlinks = true,
            linkcolor = teal,
            urlcolor  = teal,
            citecolor = teal]{hyperref}

\title{Quantifying metadata relevance to network block structure using description length}

\author[1,2,*]{Lena Mangold}
\author[1,2]{Camille Roth}
\affil[1]{Centre d'Analyse et de Math\'ematique Sociales (CNRS/EHESS), 54 Bd Raspail, 75006 Paris, France, }
\affil[2]{Computational Social Science Team, Centre Marc Bloch (CNRS/MEAE), \hbox{Friedrichstr.} 191, 10117 Berlin, Germany}

\affil[*]{lena.mangold@cnrs.fr}

\newcommand{\bfA}[0]{\mathbf{A}}
\newcommand{\bfe}[0]{\mathbf{e}}
\newcommand{\bfb}[0]{\mathbf{b}}
\newcommand{\bfk}[0]{\mathbf{k}}
\newcommand{\bfn}[0]{\mathbf{n}}

\newcommand{\ein}{e_{\mathrm{in}}}
\newcommand{\eout}{e_{\mathrm{out}}}

\newcommand{\bfem}[0]{\mathbf{e}'}
\newcommand{\bfbm}[0]{\mathbf{d}}
\newcommand{\bfnm}[0]{\mathbf{n}'}

\newcommand{\bfbo}[0]{\mathbf{b}_{\mathrm{opt}}}

\newcommand{\sopt}[0]{\Sigma_{\mathrm{opt}}^{m}}
\newcommand{\sm}[0]{\Sigma_{d}^{m}}
\newcommand{\srand}[0]{\Sigma_*^{m}}
\newcommand{\gm}[0]{\gamma_{d}^m}
\newcommand{\gdc}[0]{\gamma_{d}^{\mathrm{DC}}}
\newcommand{\gndc}[0]{\gamma_{d}^{\mathrm{NDC}}}
\newcommand{\gpp}[0]{\gamma_{d}^{\mathrm{PP}}}
\newcommand{\bfg}[0]{\boldsymbol{\gamma}_{d}}

\def\multiset#1#2{\ensuremath{\left(\kern-.3em\left(\genfrac{}{}{0pt}{}{#1}{#2}\right)\kern-.3em\right)}}

\begin{document}

\flushbottom
\maketitle

\begin{abstract}
Network analysis is often enriched by including an examination of node metadata. In the context of understanding the mesoscale of networks it is often assumed that node groups based on metadata and node groups based on connectivity patterns are intrinsically linked. This assumption is increasingly being challenged, whereby metadata might be entirely unrelated to structure or, similarly, multiple sets of metadata might be relevant to the structure of a network in different ways. We propose the metablox tool to quantify the relationship between a network’s node metadata and its mesoscale structure, measuring the strength of the relationship and the type of structural arrangement exhibited by the metadata. We show on a number of synthetic and empirical networks that our tool distinguishes relevant metadata and allows for this in a comparative setting, demonstrating that it can be used as part of systematic meta analyses for the comparison of networks from different domains.
\end{abstract}

\thispagestyle{empty}

\section*{Introduction}
Block structure in networks is characterised by the grouping of nodes on the basis of shared connectivity patterns \cite{lorrain_structural_1971, white_social_1976}. Such networks can be generated by Stochastic block models (SBMs) \cite{holland_stochastic_1983} which -- in turn -- can be used as baseline models to infer block structure from observed networks. The latter is becoming increasingly popular due to a number of considerable advances in the development of SBM variants with closer resemblance of real network structure \cite{karrer_stochastic_2011}, the introduction of flexible, nonparametric inference approaches \cite{peixoto_nonparametric_2017}, and increasingly efficient inference algorithms \cite{peixoto_efficient_2014, peixoto_merge_2020}. Blocks may take the shape of commonly studied mesoscale structures, such as assortative communities or internally cohesive clusters. However, other structural arrangements on the mesoscale, such as core-periphery structures, disassortative (bipartite) structures as well as (nested) combinations of the above, are also possible, owing to the relatively general definition of similarity of the SBM. It is often assumed that blocks -- whichever specific structural arrangement they may have -- correspond to a latent `meaning', \hbox{i.e.} some external similarity exhibited by the nodes that has made it more likely for them to connect to other nodes in the network in a similar way, or vice versa. In practice, this `meaning' is often attributed to additional information on the network nodes, which we call \emph{metadata}. In the literature, node attributes available as part of network analyses have sometimes been assumed to be intrinsically linked to the network's structure, an assumption that -- as has been demonstrated on multiple occasions \cite{newman_structure_2016, hric_network_2016, peel_ground_2017} -- cannot be readily made. 

Imagine a set of users who interact with each other on some social media platform and for whom -- in an idealised scenario -- some metadata is known to us: for each user we know their preference for one of two political parties. We can construct a network that represents the users' conversation around some political topic, by placing an edge between user nodes who interacted within the context of the topic. Assume we observe \emph{homophily} in political leaning (called \emph{assortativity} in network science): users with shared party preferences are more likely to endorse each other's content than that of users who support other parties, something that has been observed in the literature around online interactions repeatedly \cite{adamic_political_2005, conover_political_2011, barbera_tweeting_2015}. The static `snapshot' of an interaction network between these users, which might look something like the toy network in Figure \textbf{1a}, would show an alignment of a partition of users by party preference and one according to connectivity patterns. We could also say that the \emph{generative process} of this network was, at least to some extent, likely governed by this particular set of metadata under an assortativity modeling assumption. 

\begin{figure}
\centering
\includegraphics[width=0.6\textwidth]{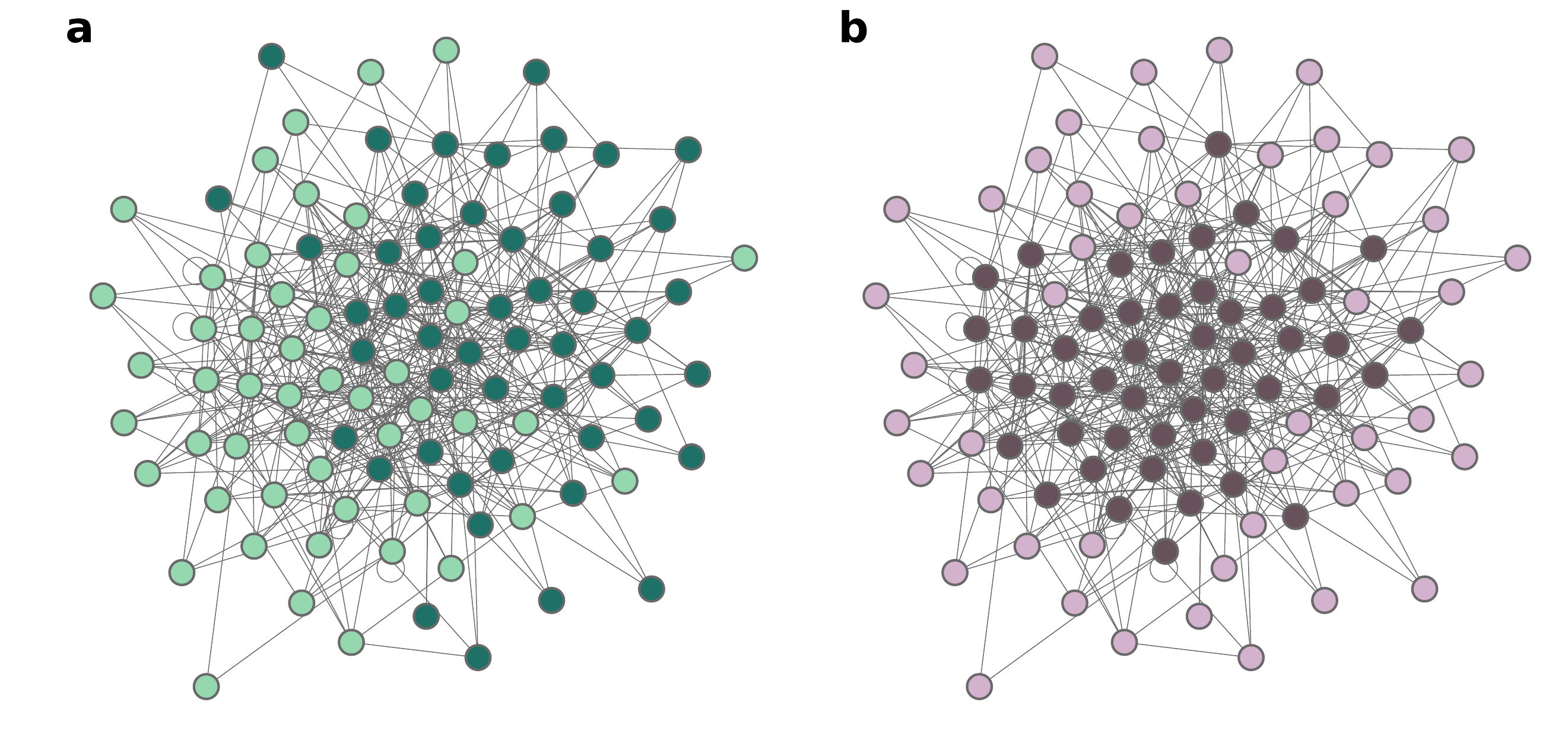}
\caption[Two partitions of a toy network.]%
{Two partitions of a toy network. \par \small \textbf{a} Example network described in main text, with nodes coloured according to their block membership in the planted bicommunity partition. \textbf{b} The same example network, with nodes coloured according to their block membership in the planted core-periphery partition. Both visualisations have been drawn using the graph-tool library \cite{peixoto_graph-tool_2014}, which is used for all network visualisations in this paper.}
\label{fig:schem_scbm}
\end{figure}

Such alignment between metadata partitions and block structure has been observed in social networks on many occasions \cite{small_structure_1974, zachary_information_1977, traud_social_2012} which may partly explain the widespread assumption of an intrinsic connection between metadata and node structure. Motivated also by a lack of knowledge on the `real' generative processes of empirical networks, node metadata has often been viewed as \emph{the} ground truth for the block structure of a network, to evaluate \cite{karrer_stochastic_2011, yang_defining_2012, chakrabort_computer_2013} or to improve \cite{yang_community_2013, bothorel_clustering_2015, binkiewicz_covariate_2017, contisciani_community_2020, fajardo_node_2022} the performance of community detection or other inference algorithms. However, a number of recent works in the complex networks community has challenged the notion of an intrinsic alignment between metadata and community structure \cite{newman_structure_2016, hric_network_2016, hric_community_2014, sweet_estimating_2018}; metadata might be entirely unrelated to structure or, similarly, multiple sets of metadata might be relevant to the structure of a network in different ways \cite{peel_ground_2017}. We can illustrate this on our example set of social media users for which we imagine a second set of node metadata: whether or not a user is an expert on the political topic that is being discussed in this particular network. Assume that in terms of structural features, the network exhibits some assortativity (of party preference) but the network also has a relatively well connected core of users -- in which there are connections even between users of different party preference -- and a loosely connected set of peripheral nodes. This structure then corresponds to the node attribute of expert (core) vs non-expert (periphery), an example of this can be seen in Figure \textbf{1b}. Overall, we thus have two sets of metadata that are both linked to the network structure while exhibiting different structural arrangements.

Besides emphasising the possible existence of multiple relevant metadata, this example can also be understood in the context of the diversity of likely node partitions exhibited by real networks: when focusing purely on the connectivity patterns in a network we can, in many cases, identify multiple, potentially qualitatively different partitions that divide the nodes in a `plausible' way \cite{peixoto_revealing_2021}. This, in turn, accentuates the notion that multiple sets of metadata can be related to the network structure, even if they divide the network's nodes in very different ways. It has been demonstrated, for example, that we can generate synthetic networks whose mesoscale -- like the one in our toy example -- is similarly well explained by (a) a division into two assortative communities and (b) a division into a well-connected core and a sparsely connected periphery \cite{mangold_generative_2023}. By fitting an SBM to such a network and sampling from the posterior distribution of partitions, we are likely to find two locally optimal partitions -- a bicommunity and a core-periphery partition -- which may be aligned with different sets of metadata. In other words: not only might multiple sets of metadata be relevant to the network structure in general, but they might be relevant in structurally very different ways.

To add to the complication, we can extend our social media thought experiment by considering the `reversed' situation: instead of one network representing a political discussion with multiple sets of metadata, we imagine multiple networks with node attributes of one particular type. Firstly, we imagine that we are able to construct \emph{multiple} networks for the same set of user nodes, such as networks of different interaction types (e.g. endorsement vs exchange of opinions) or snapshots recorded at different points in time. Secondly, we might have collected data on different topics from a variety of categories, leading to the construction of multiple topic-induced networks with different (but potentially overlapping) user node sets but shared type of metadata. To summarise and in all generality, we can divide our thought experiment into three scenarios worth thinking about: when studying social networks, we might be dealing with a situation with (I) a single network and multiple sets of metadata, (II) one set of metadata and multiple networks with the \emph{same} user sets, and finally (III) one set of metadata and multiple networks in the same \emph{context}, but with possibly different node sets.

Having outlined three types of scenarios, we come back to the questions we might be asking to investigate metadata-block structure relationships: 
Is there a way to measure the strength of the relationship between metadata and network block structure, to enable comparisons between sets of metadata for one given network (I) and multiple metadata-network pairs (II and III)? If a set of node attributes \emph{is} relevant to the structure of a particular network, can we quantify the particular structural arrangement at hand (assortativity vs some other structure)? Answering these questions requires rigorous, statistically grounded methods and we argue that generative models lend themselves particularly well in this context: they provide mathematical justification, help us deal with uncertainties connected to model selection and facilitate comparisons of different models and networks.

To our knowledge, we lack a framework that makes it possible, for a single given network and in a comparative setting for a collection of networks, to appraise the relative strength of the connection between various sets of metadata and various types of intermediate-scale structures. Related work that has gone furthest in this direction, and that serves as a motivation for our measure, is that by Peel at al \cite{peel_ground_2017}. The authors demonstrated convincingly that multiple ground-truth partitions can be responsible for the generation of a given network. They proposed two separate approaches, one to measure the statistical significance of the metadata-structure connection and one to explore the relationship between specific sets of metadata and the partition landscape of a network. Their first measure (which we outline later on in this work) serves as a p-value and thus answers a simple yes-no question with respect to the significance of metadata-structure relevance; their second measure is an inferential approach based on SBMs, which -- upon visual inspection of the results -- provides insights into the extent to which different sets of metadata are related to different parts of the partition landscape of a network. While these measures enable an in-depth analysis of individual networks and their metadata, they cannot be easily used for direct comparative purposes, since the strength of metadata-structure relationships cannot be measured and visual analysis is required for comparative studies of multiple sets of metadata; a direct comparison between networks is also not possible. Another relevant method is a label propagation approach to calculate a p-value for the significance of metadata and structure \cite{stanley_testing_2018} has similar problems and thus does not lend itself well for large-scale comparative meta analyses of multiple networks either.

In this work, we propose the \emph{metadata block structure exploration (metablox)} tool, which utilises microcanonical SBMs and methods from information theory to quantify the connection between node metadata and the block structure of a network. Our measure exploits the feature of the minimum description length (MDL) principle \cite{grunwald_minimum_2007}, which penalises overly complex models, and enables comparison across multiple sets of network-metadata pairs with respect to the \emph{strength} of the relevance of the metadata to the block structure and regarding the prominent \emph{type} of block structure exhibited by the metadata. We design the measure to enable such comparative analyses for the scenarios I-III discussed above.

\section*{Results}
\subsection*{General approach}
To derive our measure, we first note that a set of categorical node attributes naturally divides the nodes of a network into a partition. We will refer to this division of the nodes as the \emph{metadata partition} and denote it by the vector $\bfbm = \{d_i\}$, where $d_i$ denotes the metadata label of node $i$. The perhaps most obvious way to measure the relationship between a metadata partition and the block structure of a network would be to infer the optimal partition $\bfbo$ of the network in some way and then measure the alignment between the metadata partition and the optimal partition through a similarity measure $s=f(\bfbo, \bfbm)$, such as the Rand index \cite{rand_objective_1971}, variation of information \cite{meilua_comparing_2003}, or partition overlap \cite{peixoto_revealing_2021}. However, this approach may fail as soon as the detection method yields multiple similarly likely partitions that are not well-aligned in terms of their node groups: if the metadata partition is similar to one of the highly likely partitions that is not \emph{the} optimal one, we would not capture the metadata relevance.

\begin{figure}[h!]
    \centering
    \includegraphics[width=.6\linewidth]{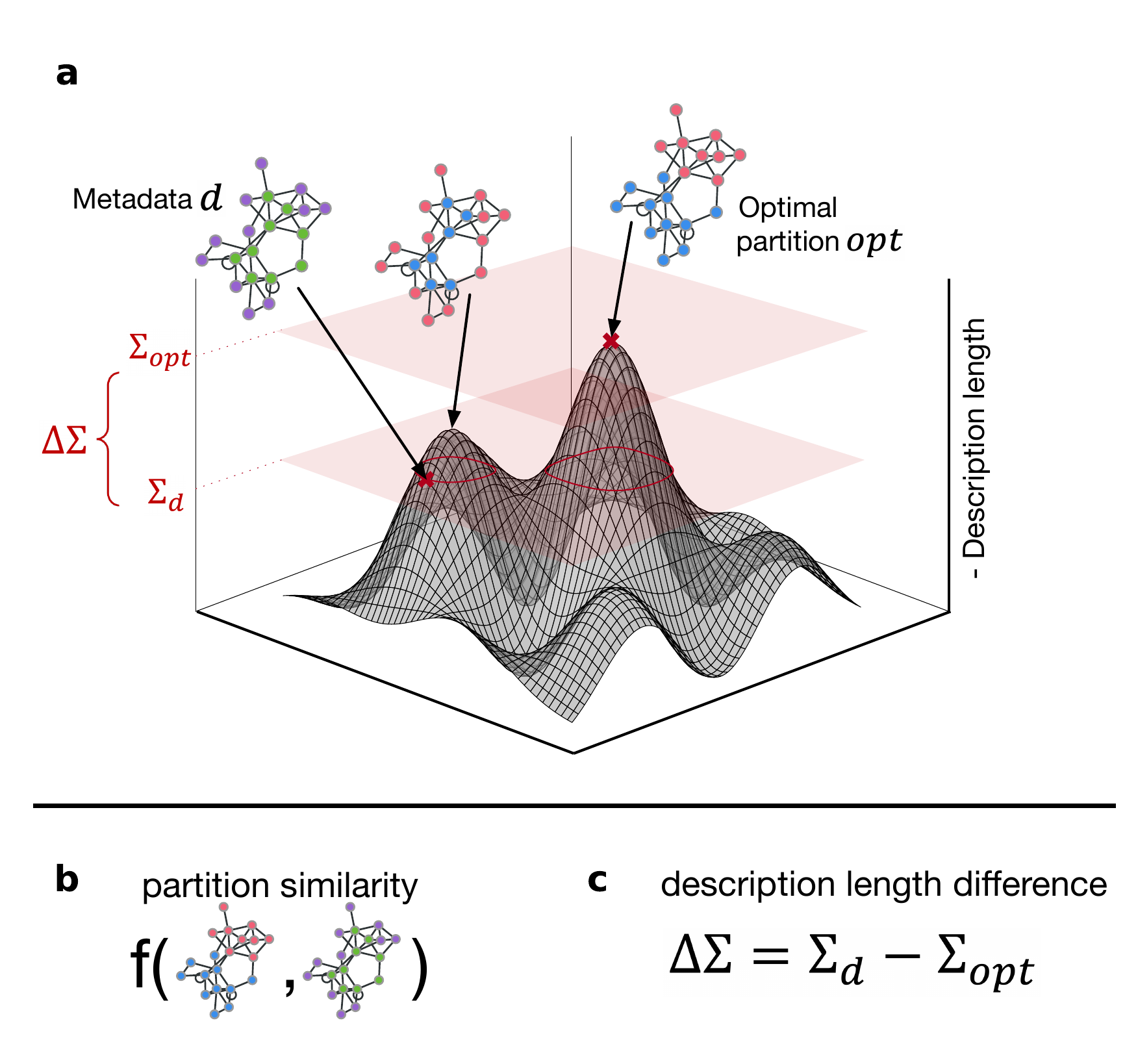}
    \caption[Schematic of a partition landscape.]%
    {Schematic of a partition landscape. \par \small 
    \textbf{a} Partition landscape \cite{peel_ground_2017, peixoto_revealing_2021} for a toy network, with the negative description length on the vertical axis, for which we have highlighted the positions of the optimal inferred partition (blue/red), a second partition (blue/red), and a metadata partition $d$ (purple/green).
    \textbf{b} Approach of measuring partition similarity directly.
    \textbf{c} Proposed metablox approach, of measuring the distance of the network's description length under the metadata partition from the network's description length under the optimal partition.}
    \label{fig:fig_schematic}
\end{figure}

Instead of directly measuring partition similarities, we take a Bayesian approach to the problem. Specifically, we note that in the Bayesian SBM framework, one generally considers the entire distribution over the possible partitions that could have been responsible for generating the network under the given SBM. With this in mind, we hypothesise that a metadata partition could -- in theory -- appear among the partitions in the posterior distribution of the SBM with some non-zero probability. Unfortunately, calculating the posterior distribution of the SBM exactly turns out to be an intractable problem and one therefore needs to resort to approximations, \hbox{e.g.} by sampling from it using an MCMC method. Since we cannot rely on finding an exact copy of our metadata partition among the samples taken from the posterior, we instead take the reverse approach: we determine where in the posterior distribution the metadata partition \emph{would} fall, were it to have been inferred by fitting an SBM to our observed network. The schematic in Figure \textbf{2a} shows an idealised partition landscape of a synthetic network (upon fitting an SBM) and illustrates the idea of the metadata being positioned more closely -- in terms of description length (\hbox{i.e.} model fit) -- to a non-optimal partition than to the optimal partition; in panels \textbf{2b} and \textbf{2c}, we show side-by-side the approach of directly measuring partition similarities described above and the approach we will be taking in this work, respectively.

\subsubsection*{Microcanonical SBMs and description length}
To lay the groundwork for this approach, we first outline the microcanonical SBM framework and its connection to the concept of description length, which we will use to quantify the metadata relevance. We focus on the so-called degree-corrected SBM \cite{karrer_stochastic_2011}, which configures an ensemble of networks with block structure, while making it possible to take into account heterogeneous degree distributions (in contrast to the traditional SBM). For a network with $N$ nodes and $B$ blocks, it is parameterised by an $N$-dimensional block membership vector $\bfb = \{b_i\}$, a degree sequence $\bfk = \{k_i\}$, and a $B \times B$ edge count matrix $\bfe = \{e_{rs}\}$. In its microcanonical form, the degree sequence and edge count matrix are specified \emph{exactly}, making it possible to count the number of possible networks that can be generated with a set of parameters, so that the likelihood $P(\bfA|\bfk,\bfe,\bfb)$, \hbox{i.e.} the probability of observing a network $\bfA$ given the SBM parameters, can be calculated as $\Omega(\bfk,\bfe,\bfb)^{-1}$, the inverse of the ensemble cardinality \cite{peixoto_entropy_2012}. When using a parametric framework to detect the optimal partition from a network, one would find the partition $\bfb$ that maximises the log-likelihood, which amounts to minimising the microcanonical entropy $S = \ln\Omega(\bfk,\bfe,\bfb)$ \cite{bianconi_entropy_2009, peixoto_entropy_2012}.

In the \emph{nonparametric} approach -- which allows for the number of model parameters to be inferred from the data --, one considers the joint distribution of the generative model, including the notion of priors on the model parameters in addition to the model likelihood. The hard constraints of the microcanonical approach imply that one does not need to sum over the remaining parameters to calculate the marginal likelihood. The joint distribution can therefore be written as $P(\bfA, \bfk, \bfe, \bfb) = P(\bfA|\bfk, \bfe, \bfb)P(\bfk|\bfe, \bfb)P(\bfe|\bfb)P(\bfb)$. This is where an information theoretical lens lets us write $P(\bfA, \bfk, \bfe, \bfb) = \mathrm{e}^{-\Sigma}$, where $\Sigma = - \ln P(\bfA|\bfk, \bfe, \bfb) - \ln P(\bfk, \bfe, \bfb)$ is called the description length of the data: the amount of information necessary to describe a network given a model plus the information required to describe the model itself, via its parameters \cite{peixoto_nonparametric_2017}. Maximising the posterior distribution is therefore equivalent to minimising the SBM description length, which -- in turn -- is the same as finding the partition which provides the most compact \emph{compression} of the network and therefore amounts to being the best model. Similarly, one can identify the more likely of two partitions $\bfb_1$ and $\bfb_2$ (under SBM variants $m_1$ and $m_2$ respectively) by comparing the network's description length for each of them, which amounts to calculating their posterior odds ratio. For $\Sigma_1<\Sigma_2$, for example, partition $\bfb_1$ (under variant $m_1$) is the more likely of the two (see Methods section for a detailed explanation). Note that we use description length as the basis for model selection (\hbox{i.e.} finding the best model by penalising overly complex modeling choices) since it is the natural choice when considering a posterior inference framework, but in certain circumstances other possible model selection criteria could be considered as well. In particular, we favour it over popular existing information criteria (such as AIC \cite{akaike_new_1974} and BIC \cite{schwarz_estimating_1978}), which are built on assumptions that do not hold for comparisons between the standard and degree-corrected SBM \cite{yan_model_2014}. Another commonly used model selection approach is cross-validation, which is widely used in machine learning and certain network science applications that allow for splitting the data into test and training sets (\hbox{e.g.} link prediction). While cross-validation can help reject models that are overly complex by favouring those that generalise better to unseen data, it is not inherently designed to penalise model complexity as directly as criteria like description length.

\subsubsection*{Measuring metadata relevance}
To tie in these concepts with our metablox measure, we realise that one can simply count the nodes in each \emph{metadata block} (\hbox{i.e.} those that share the same metadata category), and the edges within and between them, and arrive at quantities that are equivalent to the parameters of the microcanonical SBM. Given the metadata partition by $\bfbm$, we denote the edge counts within and between metadata blocks by $\bfem$. By plugging the quantities derived from the metadata and from the observed network into the SBM description length calculations -- which, in the microcanonical case, can be done exactly by using combinatorics \cite{peixoto_nonparametric_2017} -- we can straightforwardly calculate $\sm$, the description length of the network under the metadata partition $\bfbm$ and a given SBM variant $m$. In this work, we use the description length calculations from existing work \cite{peixoto_nonparametric_2017}, which we provide in the Methods section. Note that a more consistent notation would be to denote a metadata partition by $\mathbf{b}'$, but we choose $\bfbm = \mathbf{b}'$, to make a clearer distinction from the inferred partitions.

Calculating the metadata description length in this way does not yet tell us anything about the fit of the metadata \emph{relative to the partition landscape} of the network. We therefore also separately fit the SBM variant $m$ to our observed network (ignoring the metadata labels), using the graph-tool library \cite{peixoto_graph-tool_2014}, and compute the description length $\sopt$ of the optimal partition. We can then simply calculate the description length difference of the two quantities, $\Delta \Sigma = \sm - \sopt$ to understand how well the metadata partition fits the given network, compared to the optimal partition: if $\Delta \Sigma$ is close to zero we conclude that the metadata partition is strongly relevant to the block structure; if $\Delta \Sigma \gg 0$, the metadata partition fits the network much less well than the optimal partition, under the SBM variant $m$. Note that for better readability, we dropped the superscript $m$ for $\Delta \Sigma = \Delta \Sigma^m$.

\subsubsection*{Statistical significance and comparability}
There are two major caveats of quantifying the metadata relevance by simply using $\Delta \Sigma$: description length increases with growing networks size, which means that simply taking the absolute description length difference does not enable inter-network comparisons and we therefore need to normalise the measure in some way. Furthermore, we need to include a notion of statistical significance, to ensure that the fit of the metadata partition is, in fact, better than a partition we would find at random. We address both caveats at once by turning our measure into a ratio and including a distinction from randomness as follows:
additionally to $\sm$ we calculate the description lengths of the same network under an SBM with multiple sets of \emph{randomised} metadata. Each time we randomise the nodes' metadata labels, we can plug in the resulting quantities $\bfbm_*$ and $\bfe_*$ in the description length formulas again. This approach is inspired by the blockmodel entropy significance test (BESTest) \cite{peel_ground_2017}, which produces a metadata p-value, \hbox{i.e.} the probability that a randomised version of the observed metadata describes the network better than (or equally well as) the observed metadata. The BESTest p-value is calculated by generating a large set of randomised metadata partitions (whereby the number of elements in each metadata category is fixed), computing the SBM entropy under each of the randomised partitions, and finally comparing the SBM entropy under the observed metadata partition to that of the randomised partitions. However, SBM entropy decreases for an increasing number of groups which greatly complicates direct comparison of metadata models (\hbox{i.e.} different metadata partitions and/or different SBM variants). Description length enables model selection and can thus be used to compare different partitions, whether induced by metadata categories or inferred based on connectivity patterns. 

Instead of calculating a p-value from distributions of description lengths, we identify the \emph{randomised metadata description length} $\srand$, which -- for a significance level of $\alpha=0.01$ -- is equal to the first percentile of the description length distribution of the randomised metadata partitions under SBM variant $m$. The choice of $\alpha$ is, to some extent, arbitrary; here, we make the choice based on conventions in measuring statistical significance, according to which $\alpha=0.01$ denotes strong evidence against the null hypothesis, \hbox{i.e.} that the network is described equally well by randomised metadata. We emphasise that other choices can be made when our measure is used, as the required level of significance may depend on the specific context. In our measure, we take $\srand$ to be the maximum description length for which we would still consider the description length of the \emph{observed} metadata to be relevant, and we use to normalise $\Delta \Sigma$ by $\Delta \Sigma_* = \srand-\sopt$.

\subsubsection*{Probing structural arrangement}
Before we put these components together, we note that we can use the above approach to formulate a measure that not only quantifies metadata relevance but that also helps us probe for specific \emph{types of structural arrangements} of the metadata blocks: Since we can follow the above procedure for any SBM variant for which there is a microcanonical formulation with a straightforward description length calculation, we can directly plug in the structure-specific SBM variant of interest.
In this work, we focus on three SBM variants that are specific to three different types of block structures, which we briefly outline in the following. In the `standard' SBM (\hbox{i.e.} non-degree-corrected, NDC), the placement of an edge between two nodes depends solely on the block membership of each node and on the probability of two nodes from the two blocks being connected. Similar to the random graph model, one major drawback of this model is that the generative process produces networks that have blocks within which node degrees are Poisson distributed. To overcome this caveat, that makes the model unlike many real networks whose node degrees often have power-law degree distributions, the degree-corrected variant (DC) -- which we introduced above to motivate out approach -- was proposed \cite{karrer_stochastic_2011}. In this variant, edge placement depends on node degrees as well as block membership, thus accounting for heterogeneous degree distributions. A third variant we include in our analysis is the assortative `planted partition' SBM (PP), which -- as part of the generative process -- assumes assortativity \cite{zhang_statistical_2020}. The description length formulations for the three variants used in our measure are based on the work in Refs. \cite{peixoto_entropy_2012, peixoto_nonparametric_2017, zhang_statistical_2020}, as detailed in the Methods section. 

Note that we are focusing here on undirected simple graphs, so that any network for which our measure is used has to be treated as such. An extension of our measure to directed networks and/or networks with multi-edges is straightforward and should be considered as part of future research. 

\subsubsection*{Measure}
We finally put together these components to create the metadata block structure exploration (metablox) tool, which produces the vector $\bfg$ for a metadata partition $d$, where each element represents the metadata relevance under a different SBM variant $m$. It can consist of as many elements as we include SBM variants to probe for different structural arrangements. Here, we will work with the two variants used in the motivating example, DC and PP as well as the non-degree-corrected SBM (NDC), thus $\bfg$ is a vector with three elements. For a metadata partition $d$, the metablox vector $\boldsymbol{\gamma}_d$ thus consists of elements
\begin{equation}
\label{eq:gamma}
    \gm = \frac{\Delta \Sigma^m}{\Delta \Sigma^m_*} = \frac{\sm-\sopt}{\srand-\sopt}
\end{equation}
for each SBM variant $m$. In summary, $\sm$ is the description length of metadata partition $d$ under model $m$, $\srand$ is the randomised metadata description length, and $\sopt$ is the description length of the optimal partition of the network. Recall that the difference in description length between two models is equal to the log of the posterior odds ratio of the two models. The numerator of $\gm$ thus measures how much more likely the optimal partition is compared to the metadata partition under model $m$. Similarly, the denominator represents how much more likely the optimal partition is to the randomised metadata partitions under all models. The total measure is the former normalised by the latter.

With this definition, each element can then be interpreted as follows: for $\gm\geq1$ we say that this set of metadata is not relevant to the structure under model $m$, since more than $1\%$ of the randomised metadata partitions compress the network more efficiently than our observed metadata, under the given model. For $\gm < 1$, the metadata is relevant to the structure under the given model, and the closer $\gm$ is to $0$, the stronger the relevance of the set of metadata to the block structure and the closer the structural arrangement of the metadata to the typical type of structure generated under the given model $m$. Another way of interpreting $\gm$ is: relative to the best compression we can find, how efficient is the compression of the observed metadata $d$ under $m$ compared to partitioning the network by randomised versions of the metadata. 

\subsubsection*{\label{sec:limitations}Limitations and network compressibility}
One limitation of our approach is that the inequality $\gamma^{m_1}_d < \gamma^{m_2}_d$ for a network with some metadata partition $d$ merely indicates that the metadata $d$ is relatively closer to the best partition under model $m_1$ than to the best partition under model $m_2$. It does not necessarily mean that the block structure of our network as a whole is optimally explained by $m_1$: since our measure is a ratio of description length differences, it is possible that in fact $\Sigma_{\text{opt}}^{m_1} > \Sigma_{\text{opt}}^{m_2}$ or that the model $m_2$ provides a better explanation for the network overall. This limitation arises from the comparative nature of the measure, which inherently focuses on relative rather than absolute fit to enable comparisons across metadata and networks.

Another limitation is related to the case of inter-network comparison, in particular in the context of different network topology. While metablox is robust with respect to the number of nodes in a network (see Methods section), certain aspects of a network's topology do affect the measured metadata relevance. Let us assume that we are comparing two node attributed networks $A$ and $A'$, with the signal in the community structure of network $A$ being significantly stronger than that of network $A'$ (\hbox{i.e.} the communities of network $A$ are more clearly separated). All other things being equal -- including the level of alignment of the metadata partition with the optimal detected partition -- we will have $\gamma < \gamma'$. Since $\sopt$ decreases when the network becomes more compressible due to a stronger signal in the block structure, $\Delta \Sigma*$ increases (since $\srand$ does not decrease with $\sopt$) and does so more quickly than $\Delta \Sigma$. This apparent conflation of block structure signal strength and metadata block-structure relevance points to a question around the definition of this `relevance', which in turn relates back to our earlier point of going beyond the direct measurement of partition alignment. We argue that the concept of measuring the relevance of metadata to block structure in fact \emph{should} include the notion of block structure strength: we do not want to retrieve the same $\gamma$ for the described networks $A$ and $A'$ merely because there is alignment of metadata partition and optimal partition. Instead, the relevance of the metadata should be quantified as stronger (\hbox{i.e.,} a better $\gamma$) when the network exhibits more significant block structure.

The elements of the measure that lead to these limitations are part of the design that enables comparative work and what enables the measure to capture metadata-structure relationships in the intended way. However, there might be reasons for a researcher using this measure to require insights -- alongside the metablox vector $\bfg$ -- into the absolute fit of the optimal partition, which acts as the main point of comparison in the partition landscape, or to disentangle block structure signal and metadata partition correlation. To account for this, the edge compression $c^{m}=\sopt/E$ can be referred to as a second dimension in our measure, which quantifies the absolute compression of the optimal partition per network edge. It can serve as a point of reference for the structure of the optimal partition and help us understand how much of $\gamma$ is explained by correlation vs compressibility of the network under a given model. In the Methods section we include an analysis of the sensitivity of our measure to network topology and the role of the second dimension.

\subsection*{Simulations}

We now summarise the metablox pipeline and state the choices we make for any calculation of $\bfg$ in the remainder of this work: We calculate $\sm$ by plugging the required quantities, as obtained by the metadata partition $\bfbm$, into the description length formulas for SBM variant $m$. For $\sopt$, we use the graph-tool library to fit variant $m$ to the observed network and find the partition $\bfbo^m$ that minimises the description length, refining the result of the agglomerative algorithm by running $1000$ sweeps of the merge-flip MCMC \cite{peixoto_merge_2020}. We calculate $\srand$ by performing $500$ permutations of the metadata labels, calculating the description length of the network under each of them, and finding the description length for which only $1\%$ of permutations have a lower description length.

\subsubsection*{Synthetic network}

To illustrate the metablox measure, we generate a network with multiple, non-aligned, plausible explanations in terms of its block structure, create multiple sets of synthetic metadata and calculate metablox on two dimensions: under a degree-corrected (DC) and a planted partition (PP) SBM. More specifically, we generate the network with the Stochastic cross block model (SCBM) \cite{mangold_generative_2023}, which facilitates the generation of such networks with `ambiguous' mesoscales structures, by `planting' (similar to the planted partition model \cite{condon_algorithms_2001}) multiple coexisting partitions. We use the SCBM to generate a network with $N=100$ nodes, $B=2$ blocks and expected degree $k=10$. We plant two coexisting partitions, so that the network exhibits bicommunity (BC) structure (\hbox{i.e.} two assortative communities) as well as core-periphery (CP) structure (see the Methods section for details on how this network was constructed). Figure 1 shows this network in two separate visualisations, with nodes painted according to their block membership in the BC partition in Figure 1a and according to their block membership in the CP partition in Figure 1b. We also generate multiple sets of synthetic metadata, in a way such that the labels align with the block structure of the network to varying degrees, similar to related work \cite{peel_ground_2017}. We correlate half of the synthetic metadata with the BC partition and the other half with the CP partition, with varying level of correlation $\rho$. The metadata label of each node is equal to the node's block assignment of the planted structure with probability $(1+\rho)/2$. We increase $\rho$ from $0$ to $1$ at steps of $0.01$, yielding a total of $202$ sets of metadata, half of which being `bicommunity-like' and the other half being `core-periphery-like'.

In Figure \textbf{3}, each dot represents one set of metadata $d$ for which we are plotting $\gamma_d^{DC}$ and $\gamma_d^{PP}$ for increasing correlation $\rho$, for the bicommunity-like metadata (Figure \textbf{3a}) and the core-periphery-like metadata (Figure \textbf{3b}). For Figure \textbf{3} only, we additionally calculate the BESTest p-value to illustrate the contribution of metablox compared to the most similar existing measure \cite{peel_ground_2017}.

\begin{figure*}
\centering
\includegraphics[width=0.95\textwidth]{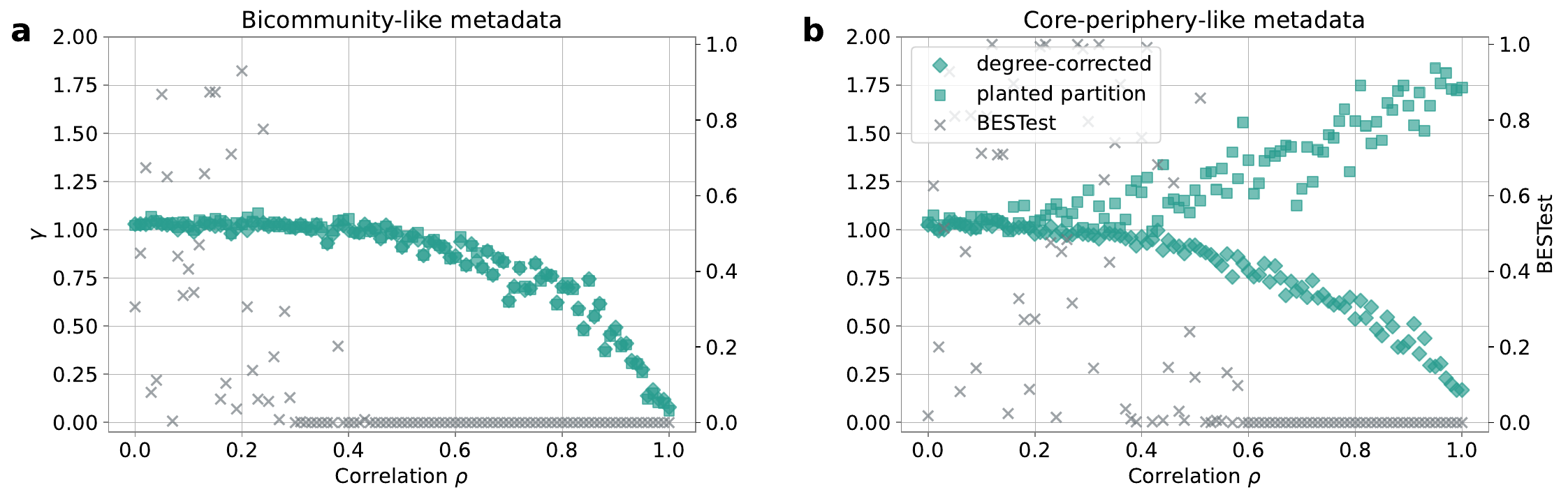}
    \caption[Metablox and BESTest values for a synthetic network.]%
    {Metablox and BESTest values for a synthetic network. \par \small 
    Metablox values for the degree-corrected and planted partition SBM, and BESTest values \cite{peel_ground_2017} under the degree-corrected SBM, for a synthetic network with multiple sets of metadata. 
    \textbf{a} Values for the sets of bicommunity-like metadata.
    \textbf{b} Values for the sets of core-periphery-like metadata. Both panels show an increasing correlation $\rho$ between metadata and block structure on the x-axis.}
\label{fig:compare_bestest}
\end{figure*}

For the bicommunity-like metadata, we can see that both dimensions of $\gamma$ are a decreasing function of the correlation $\rho$: the more strongly correlated the metadata, the higher the relevance assigned to it by the metablox measure. As expected, we cannot make a distinction between the general (DC) and assortative (PP) SBM variant, since -- by design -- the metadata in this case is similar to the block structure that was created under an assortativity assumption. For core-periphery-like metadata, however, $\gdc$ decreases similarly to the previous case, while $\gpp$ is now an increasing function of $\rho$, suggesting that -- as we know is true -- if we were to assume that this metadata was responsible for generating this network, assortativity would be much less likely to explain its structure.

For comparison, we observe that the BESTest values become non-distinguishable once the correlation $\rho$ has crossed a certain threshold. When the SBM entropy under a given metadata partition is lower than the SBM entropy under \emph{all} sets of randomised metadata partitions then the p-value is $1/n_{\mathrm{p}}$ (where $n_{\mathrm{p}}$ is the number of randomised metadata partitions) and we cannot compare this metadata to another set of metadata for which this is also true. In our example, BESTest does not allow us to compare a metadata partition that is nearly perfectly correlated with the planted block structure, to one for which $\rho=0.7$, in both the bicommunity and core-periphery case. For comparison, the metablox values for DC cross the significance line $\gamma=1$ at a similar point as BESTest becomes significant, for both the bicommunity and core-periphery-like metadata. However, it proceeds in the shape of a continuous decreasing function with a minimum of $0$ at $\rho=1$, which is what we would expect as the metadata partition is equal to the planted block structure at this point. The first contribution of metablox is therefore that, by comparing the metadata partition to the \emph{optimal} partition (while still taking into account statistical significance), we are able to make direct comparisons between metadata sets. The second contribution is illustrated by the fact that BESTest does not tell us anything about likely structural arrangements of the metadata, while the difference between the trajectory of $\gpp$ on the left and right figure demonstrate the way in which metablox does allow for this.

We also apply the metablox measure to three real-world networks, in each case referring to one of the three scenarios (I-III) introduced above. As outlined previously, our measure can be extended to include any number of structure-specific SBMs. For our applications to real networks, we focus on the non-degree-corrected (NDC), the degree-corrected (DC), and the assortative `planted partition' (PP) SBM. This means that we calculate a three-dimensional metablox vector for every network-metadata pair.

\begin{figure}
\centering
\includegraphics[width=1\textwidth]{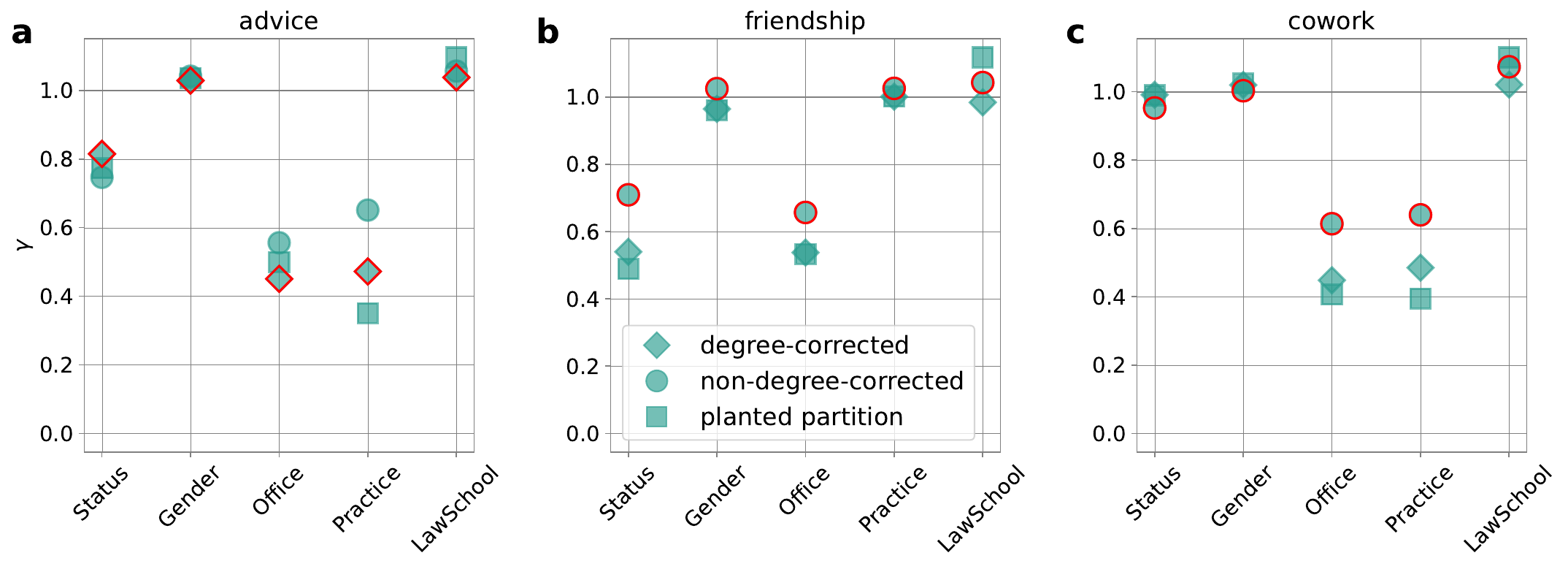}
    \caption[Metablox values for multiple metadata on law firm networks.]%
    {Metablox values for multiple metadata on law firm networks. \par \small 
    Degree-corrected, non-degree-corrected and planted partition dimensions of the metablox vector for each of five sets of metadata (status, gender, office, type of law practised, and law school attended) for three networks of employees of a law firm \cite{lazega_collegial_2001}.
    \textbf{a} Advice network.
    \textbf{b} Friendship network.
    \textbf{c} Coworking network. 
    On each figure, the SBM variant that gives the lowest edge compression for the network is highlighted in red.}
\label{fig:law}
\end{figure}

\subsubsection*{Application: Law firm}
We first demonstrate our method on a number of networks from the Lazega law firm networks collection \cite{lazega_collegial_2001}. In this collection, there are three different networks among the same set of nodes, in which edges represents different types of connections between the employees of a corporate law firm: coworkers, friendship, and advice. For the employees, we have five sets of node attributes - their status in the firm, gender, the office in which they work, which type of law they practice, and which law school they attended. Figure \textbf{4} shows the metablox results for the three networks, each on a separate figure. On each figure, we show the results for the respective network and compare the five different metadata partitions. Each figure thus represents a case of scenario I, while we can also compare network-metadata pairs which share a type of node attribute (scenario II). Note that we have highlighted the SBM variant which gave the best per edge compression with a red marker. We can see some considerable variation in how the sets of metadata are related to the networks' block structure under each of the SBM variants. Firstly, we observe that the strongest metadata-block structure relevance can be observed between the type of law practiced by an employee and the block structure in the advice network, under an assortative SBM assumption. This implies that employees are more likely to seek advice from colleagues who practice the same type of law as themselves. The same set of metadata is also strongly relevant in the cowork network. However, in the friendship network the law practiced by an employee does not seem relevant to the formation of blocks under any of the SBM variants. Employees' status, on the other hand, is the most strongly relevant metadata out of all attributes we have for the friendship network, under PP and DC. This indicates that a division of the network's nodes into the two available categories (partner and associate) in this set of metadata is more strongly aligned with a plausible partition according to shared connectivity patterns in the network, under the modelling assumptions of both PP and DC. In other words, the nodes that share a metadata attribute (partner or associate) are not only likely to share connectivity patterns independent of their node degrees (DC), they are also more likely to connect to employees with whom they share the status than with others (PP). Status matters less in the advice network (albeit still significantly relevant) and is essentially irrelevant in the cowork network. The office in which employees were located is the only set of metadata that is highly relevant for all three networks, indicating that the physical location of employees' plays a large role in determining connections. Both employees' gender and the law school they attended are not related to the block structure under any SBM variant and therefore appear to be essentially irrelevant in the tie generation process. In this example, we have been able to both compare different sets of metadata for a given network (scenario I) as well as different networks with the same node set and shared metadata (scenario II).

Peel at al. \cite{peel_ground_2017} used the same networks to demonstrate both of their methods, which we outlined above. Using the BESTest significance test, they reached similar conclusions for the network-metadata pairs. Using a second method -- which admittedly relies on visual interpretation and does not explicitly serve to compare the strength of metadata relevance -- the authors concluded that for the friendship network, the law school attended by employees was more strongly structure-relevant than the office in which employees are located. This finding is the opposite of that given by metablox. While the methods proposed by the above authors can provide insights into the significance of metadata relevance and the quality of the relationship for a given network-metadata pair, they do not enable a \emph{direct quantification} of the strength of the relationship and of the likely prominent structure. As a result, they also do not enable a direct comparison of different networks (either made up of different or the same node sets), something our measure is designed to do as shown in the following paragraphs.

\subsubsection*{Application: Impact investing}
We continue with a second example for scenario II, in which we investigate the metadata-block structure relationship in Twitter/X retweet networks of users discussing the topic of `impact investing' \cite{chiapello2020social}. In October 2021 we exhaustively collected tweets mentioning any term belonging to a small subset of hashtags directly relevant to this topic. This amounted to 1.89m tweets published by around 300k accounts since 2007. We deemed users who posted less than a tweet a month on average or who never posted more than one year apart to be insufficiently active in that field and filtered them, leading to a core set of around 16k users. We subsequently inferred likely home countries from the geolocation voluntarily shared by users in their profile description. Most users appeared to come from an English-speaking country: countries with at least 1,000 active accounts include only the US, Canada, Australia and the United Kingdom, covering 72\% of the data. Focusing on these country users, we eventually study 13 static networks representing each a snapshot for each year between 2009 and 2021 (we excluded 2007 and 2008 which yielded trivial networks) \cite{data_impact_investing}. 

\begin{figure}
\centering
\includegraphics[width=1.\textwidth]{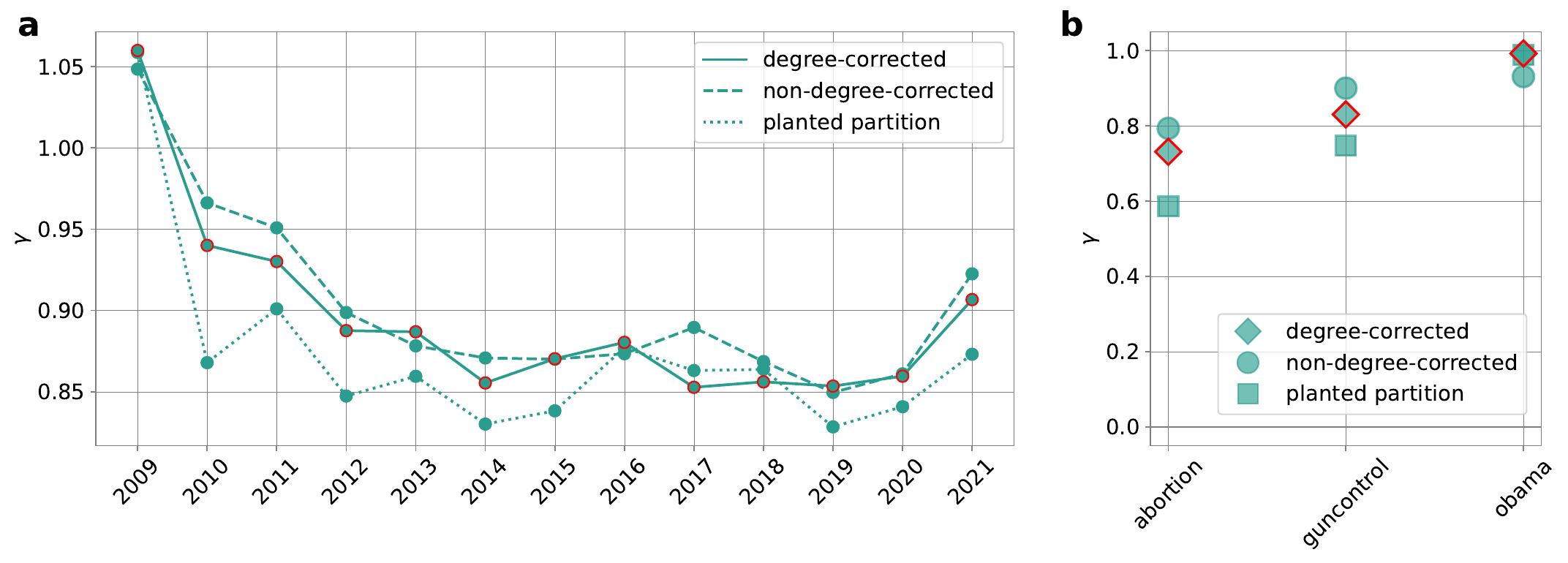}
    \caption[Metablox values for multiple Twitter/X networks.]%
    {Metablox values for multiple Twitter/X networks. \par \small 
    Degree-corrected, non-degree-corrected and planted partition dimensions of the metablox vector for various Twitter/X networks.
    \textbf{a} Static snapshots, representing a non-overlapping one year period each, of a Twitter/X retweet network among users discussing the topic of impact investing, with user location (country) as shared metadata (scenario II).
    \textbf{b} Three Twitter/X interaction networks \cite{garimella_political_2018, hohmann_quantifying_2023} with shared metadata representing the users' political stance (liberal vs conservative) (scenario III).
    The SBM variant that gives the lowest edge compression for each network is highlighted in red.}
\label{fig:twitter}
\end{figure}

This time, the edges in the different networks represent the same type of interaction at different moments in time. The metadata here is the country which the user declares being located in and which we consider to be fixed over time, being inferred at the moment of data collection in 2021. We calculated the three metablox dimensions for each network snapshot and also highlighted, for each network, the SBM variant that provided the best compression per edge for its optimal partition. In Figure \textbf{5a}, we can see that the country in which a user is located starts being relevant to the network's block structure in 2010, under all SBM variants but most strongly under PP. In the next five years (until 2015) the metadata becomes more strongly relevant in general and assortativity remains the prominent arrangement of the metadata. In 2016, there seems to be a change in the structural organisation of the network in relation to the metadata: the country in which users are located remains relevant to the block structure but in 2017 and 2018, PP is replaced by DC as the best fit for the metadata, before it swaps again starting in 2019 up until the last year of data collection. It is worth noting that for each year, DC is the SBM variant for which the optimal inferred partition provided the best compression. Overall, the quantification of the relevance of the metadata afforded by metablox enables us to observe changes over time that should serve as a starting point to `zoom into' the networks for those particular events to gain a deeper understanding of the relationship between the users' location and their tweeting behaviour: the change from 2009 to 2010 when the country user location becomes significant, the years 2014 and 2019 when the metadata is most strongly relevant under PP; and the years 2016-2018, in which the networks seem to exhibit a change in the metadata-structure relationship.

\subsubsection*{Application: Twitter/X}
In our final example, we proceed with another set of networks created from Twitter/X, this time demonstrating metablox for networks in the same context and not necessarily with the same set of nodes (scenario III). In particular, we use data from three debates on political topics in the US, for which we are studying the metadata-block structure relationship for the same type of node attributes. The tweets from which these networks were created were collected between 2015 and 2016, and are based on conversations on abortion, Obamacare, and gun control \cite{garimella_political_2018, hohmann_quantifying_2023}. In all three networks, the available node metadata reflects two categories of political orientation as either liberal or conservative. The liberal-conservative opinion categories are based on a continuous score between -1 and +1 that were calculated \cite{hohmann_quantifying_2023} based on a method that uses URLs shared by Twitter accounts and the categorisation of websites behind these URLs on https://mediabiasfactcheck.com \cite{cinelli_echo_2021}. The two categories used as node metadata are based on users below and above a `neutral' score of $0$. Note that we do not know to what extent there is an overlap between the set of users participating in the debates represented by the three networks, neither do we care; here, the subject of comparison is the topic that is discussed in each network and to what extent the political stance of users is related to the block structure for each topic.

In Figure \textbf{5b}, we plot the metablox vectors for these interaction networks with binary political orientation metadata. We observe that for the topics of gun control and abortion, the metadata partition into liberals and conservatives is relevant under all variants but most strongly under PP, while DC provides -- again -- the best overall fit. For the obamacare network, the metadata is barely relevant under NDC and not significantly relevant under the other variants. We can conclude that for gun control and abortion, the metadata partitions were likely to be at least somewhat related to the edge generating process and that assortativity is the prominent structural arrangement of the metadata in those cases. Our findings seem equivalent to previous findings on polarisation levels in these networks, which classified all three networks to be polarised to some extent with the Obamacare network being the least polarised \cite{hohmann_quantifying_2023}. Similar to our work, their polarisation measure also considers both network structure as well as a metadata dimension. However, their measure considers continuous node attributes -- specifically political ideology on a continuous left-right scale -- and specialises on measuring polarisation rather than the more general approach we take here. So although our measure has a different purpose, it might be able to pick up specific network properties (such as polarisation or fragmentation) while also being general enough to enable broader comparisons. 

\section*{Discussion}
In this paper, we have introduced a measure for probing the relationship between network metadata and its structural organisation. Our metablox pipeline, which produces the vector $\boldsymbol{\gamma}$, is designed to provide insights into the relevance of metadata to a network's block structure and the likely structural arrangement in various scenarios: for the comparison of multiple sets of metadata for one given network (scenario I) and to compare the metadata-block structure relationships for entire collections of networks that share the same type of metadata, for networks that have the same node set (scenario II) and for networks within the same context (scenario III).

We have demonstrated the metablox measure on a synthetic network and applied it to a number of real networks, including the Lazega law firm networks and networks representing different social media debates. The results reveal variations in the relevance of metadata partitions to block structure, and the likely structural arrangement, therefore providing insights into the underlying dynamics of these networks. By covering examples from the three scenarios (I-III), we have demonstrated that our measure allows for a comprehensive inter- and intra-network comparison, enabling researchers to quickly identify networks where specific metadata partitions are closely related to structure or where certain structural arrangements are likely or unlikely under the metadata. 

We have discussed a number of limitations of the measure, related to the specific design of metablox as a comparative measure, and proposed ways to address these. Additional limitations are connected with the specific type of node attributes for which relevance can be measured. In its current framework, metablox can only quantify the relevance of non-overlapping categorical metadata, while overlapping, real-valued or multidimensional metadata cannot be evaluated. For the case of overlapping metadata, it is conceivable that an overlapping SBM variant may be used. For real-valued and multidimensional metadata, the options are less clear. We therefore see a possible research direction in an alternative version of the metablox tool for real-valued metadata. While this clearly requires a number of non-trivial decisions with respect to suitable generative models, providing an equivalent measure for non-discrete metadata seems a valuable extension.

Other potential research direction evolve around the particular type of structural arrangements of metadata categories. Currently, we focus on degree-corrected (DC), non-degree-corrected (NDC), and planted partition (PP) stochastic block models. However, networks often exhibit more complex structural arrangements beyond these models and so might their metadata. Future work could explore the integration of other SBM variants that are tailored to specific structural motifs, such as core-periphery structures, bipartite structures or nested patterns. The current measure has been implemented for undirected graphs, but extensions to more complex network structures such as directed graphs are straightforward and should also be considered as part of future research. 

Our measure has the potential to serve as a tool for conducting large-scale comparisons of collections of network-metadata pairs, for networks coming from a variety of research fields -- as long as categorical node metadata is available. Obvious examples are: different types of social networks, for which metadata may include a range of demographics or affiliations; biological networks such as gene regulatory networks, protein-protein interaction networks, and ecological networks with metadata related to genes, proteins, or species; economic and financial networks, such as trade networks, supply chains, and stock market networks, for which metadata may relate to industries, sectors, or companies; or networks from science of science, where networks represent collaborations and knowledge flows, and metadata may include fields of scientific research.

\section*{Methods}
\subsection*{\label{sec:description_length}Data compression \& microcanonical SBMs}
To justify the use of description length as part of the metablox formulation, we expand on the Results section in the main text and more throughly explain the minimum description length (MDL) principle and its relationship with microcanonical SBMs. MDL is a model selection criterion according to which one should favour the model that achieves the smallest compression of the data. The idea behind this is that compression is possible when we find regularities in the data which, in turn, means that we `learn' about patterns in the data \cite{grunwald_minimum_2007}. MDL is sometimes described as a formal interpretation of Occam's razor -- also known as the principle of parsimony -- which is the idea that one should try to find the explanation with the smallest number of assumptions possible. In more formal terms, the best hypothesis $H$ (\hbox{e.g.} a model with its parameters) for a data set $D$, is the one that minimises the sum $S(H) + S(D|H)$, where $S(D|H)$ is the amount of information required to describe the data $D$ when it has been encoded with the hypothesis $H$ and $S(H)$ is the amount of information necessary to describe the hypothesis itself. This demonstrates the `automatic' overfitting-prevention property of MDL, which makes it an attractive model selection criterion: with a more complex hypothesis, we need less information to describe the data given the hypothesis, but we need more information to describe the hypothesis itself. 

There is a strong relationship between MDL and Bayesian inference in general \cite{grunwald_minimum_2007} and the Bayesian interpretation of the SBM more specifically, where it was first used to infer network partitions without knowing the number of blocks in advance \cite{peixoto_parsimonious_2013}. To calculate the description length of a network under a particular model, one needs to derive the entropy of the individual components of the SBM, which is an ensemble of networks that can be generated from a set of parameters (\hbox{i.e.} the partition). In general, \emph{microcanonical} network ensembles -- for which structural constraints need to be satisfied exactly rather than on average -- can be described by their entropy $S = \ln \Omega(\theta)$, where $\Omega(\theta)$ is the total number of networks that can be generated under the given set of parameters $\theta$ \cite{bianconi_entropy_2009}. The higher the entropy of a network ensemble, the more `disordered' (or `random') is the ensemble. More concretely, let us assume that we have a network $\bfA$ generated by a model with parameter set $\theta$. $P(\bfA|\theta)$ is the probability of observing the network $A$ in an ensemble generated by the model with these parameters (\hbox{i.e.} the likelihood) and we assume that all networks occur with the same probability $P(\bfA|\theta)=\frac{1}{\Omega(\theta)}$. From this assumption, one can straightforwardly make a connection between the microcanonical entropy $S$ and the log-likelihood: $L = \ln P = - \ln \Omega(\theta) = - S$ \cite{peixoto_entropy_2012}. By minimising the entropy $S$ one could therefore find the maximum likelihood parameters, such as the most likely partition, given an observed network. While maximum likelihood methods work well in many cases, in the particular case of model selection with SBMS, maximum likelihood estimation can lead to overfitting if the number of model parameters is not fixed. For example, if the number of blocks $B$ is not known, minimising the entropy would lead to the trivial partition of every node being its own block, \hbox{i.e.} $B=N$, where $N$ is the number of nodes. Peixoto proposed the MDL principle as part of a microcanonical \emph{nonparametric} approach to SBM inference, to address the overfitting issue and enable the use of flexible priors and hyperpriors on the model parameters \cite{peixoto_parsimonious_2013, peixoto_nonparametric_2017}. Specifically, this means considering the full joint distribution of the network and the SBM model parameters as part of the inference process, rather than just the SBM likelihood, which can be interpreted as calculating the description length of the network under the given model, via its parameters.

We recall that for a network $A$ of size $N$ with $E$ edges, we denote the degree of a node $i$ by $k_i$, and we use the following notation for the parameters of a microcanonical SBM with $B$ blocks. The block assignments of the nodes is denoted by the vector $\bfb = \{b_i\}$ of length $N$, the $B \times B$ matrix $\bfe = \{e_{rs}\}$ represents the edge counts within and between two blocks $r$ and $s$ (with twice the number of edges on the diagonal, as is convention), and the $B$-dimensional vector $\bfn = \{n_r\}$ describes the number of nodes in each block $r$. The full joint distribution of the degree-corrected SBM is, as introduced above, given by $P(\bfA, \bfk, \bfe, \bfb) = P(\bfA|\bfk, \bfe, \bfb)P(\bfk|\bfe, \bfb)P(\bfe|\bfb)P(\bfb)$. In terms of the generative process of this model, this means that one first samples a partition of the nodes into blocks, then samples the numbers of edges within and between the blocks, then samples the half-edges according to the node degrees, and finally connects half-edges accordingly to create the network (i.e. sampling from the networks that are possible given the partition and number of edges).

When using the nonparametric Bayesian framework for inference purposes, one can then use this formulation to maximise (or sample from) the \emph{posterior} distribution of partitions 
\begin{eqnarray}
    P(\bfb|\bfA) = \frac{P(\bfA,\bfb)}{P(\bfA)} = \frac{P(\bfA|\bfb)P(\bfb)}{P(\bfA)}.
\end{eqnarray}
This is where the microcanonical formulation helps simplify the inference problem, which allows us to write $P(\bfA|\bfb) = P(\bfA|\bfk, \bfe, \bfb)P(\bfk|\bfe, \bfb)P(\bfe|\bfb)$. The marginal likelihood is usually $P(\bfA|\bfb) = \sum_{\bfe} P(\bfA|\bfk, \bfe, \bfb)P(\bfk|\bfe, \bfb)P(\bfe|\bfb)$, but due to the `hard' constraints of the microcanonical SBM, there is only one non-zero element in this sum \cite{peixoto_parsimonious_2013}. 

It turns out that this is where we can re-introduce information theoretical interpretation. In particular, the posterior can be rewritten as
\begin{eqnarray}
    P(\bfb|\bfA) = \frac{P(\bfA|\bfk, \bfe, \bfb)P(\bfk|\bfe, \bfb)P(\bfe|\bfb)P(\bfb)}{P(\bfA)} = \frac{\mathrm{e}^{-\Sigma}}{P(\bfA)}
\end{eqnarray}
where $\Sigma = - \ln P(\bfA|\bfk, \bfe, \bfb) - \ln P(\bfk, \bfe, \bfb)$ is the description length. Equivalently to the more general introduction to description length above, the first component is the amount of information required to describe the network under the SBM and the given parameters and the second component is the amount of information needed to describe the parameters themselves. Due to this definition, finding the partition $\bfb$ that minimises the description length is equivalent to finding the partition that maximises the posterior, \hbox{i.e.} to finding the most likely partition of the network. 

The description length can be used to identify the most likely model given the observed data, upon comparing multiple competing models. One way of interpreting a `model' in this context is as a particular partition $\bfb_1$ under an SBM variant $m_1$. We can identify for which of two partitions $\bfb_1$ and $\bfb_2$ (under models $m_1$ and $m_2$ respectively) there is more evidence in the data, by calculating their posterior odds ratio
\begin{equation}
\begin{aligned}
\label{eq:modelselection}
\Lambda &= \frac{P(\bfb_1, m_1|\bfA)}{P(\bfb_2, m_2|\bfA)} \\
        &= \frac{P(\bfA|\bfk, \bfe_1, \bfb_1, m_1)P(\bfk|\bfe_1, \bfb_1, m_1)P(\bfe_1|\bfb_1, m_1)P(\bfb_1|m_1)P(m_1)}{P(\bfA|\bfk, \bfe_2, \bfb_2, m_2)P(\bfk|\bfe_2, \bfb_2, m_2)P(\bfe_2|\bfb_2, m_2)P(\bfb_2|m_2)P(m_2)} \\
        &= \mathrm{e}^{-\Delta\Sigma},
\end{aligned}
\end{equation}
where $\Delta\Sigma = \Sigma_1 - \Sigma_2$ and $\Sigma_i = - \ln P(\bfA|\bfk, \bfe_i, \bfb_i, m_i) - \ln P(\bfk, \bfe_i, \bfb_i, m_i)$ is the description length of model $i$ (\hbox{e.g.} of the network under SBM $m_i$ with partition $\bfb_i$). Here we assume that both variants are equally likely \hbox{i.e.}, $P(m_1) = P(m_2)$\cite{peixoto_nonparametric_2017}. The first component of $\Sigma_i$ is the amount of information required to describe the network under model $i$ and the given parameters; the second component is the amount of information needed to describe the parameters themselves. One can therefore identify the more likely model (in terms of the specific parameters) by calculating the description lengths of the network under each model and partition. For $\Lambda = 1$ or, equivalently, $\Sigma_1 = \Sigma_2$, the models are equally likely and for $\Lambda > 1$ ($\Sigma_1 < \Sigma_2$) model $m_1$ is more likely than model $m_2$.

\subsection*{\label{sec:dl_calculations}Description length calculation}
As described above, the description length $\Sigma$ of a network $\bfA$ under an SBM with a set of parameters $\theta$ can be directly derived from its full joint distribution, since $\Sigma = -\ln P(\bfA, \theta) = - \ln P(\bfA|\theta) - \ln P(\theta)$. Here, we detail the description length calculation for each of the SBM variants discussed in the main text, by providing their joint distributions, made up of model likelihood and priors. We will see that the microcanonical framework makes it possible to derive the likelihood and priors through combinatorics. Specifically, to be as parsimonious as possible, the priors tend to be uniform distributions over the number of possible realisations of a particular parameter under the given modelling assumptions. 

Note that in the formulas of the description length calculations, we use the notation introduced for the SBM parameters. However, in the metablox measure, the description length is not only used as part of the inference of the optimal partition but it is also calculated for the metadata partition. To calculate the metadata description length, we simply replace the parameters $\bfb$, $\bfe$, $\bfn$, $B$ by the respective metadata quantities that can be directly induced by $\bfbm$. For example, $\bfem$ is the number of edges within and between the node sets with shared node attributes, $\bfnm$ is the number of nodes that share each type of node attributes, 
and $B'$ is the number of unique node attributes in $\bfbm$.

The formulations we provide here for the likelihood and priors for the non-degree-corrected (NDC) and degree-corrected (DC) SBM are based on the work in Refs. \cite{peixoto_entropy_2012, peixoto_nonparametric_2017}, those for the assortative planted partition (PP) SBM are from Ref. \cite{zhang_statistical_2020}. 

For NDC, the only model parameters are the edge counts $e_{rs}$ between blocks $r$ and $s$ and the block assignment vector $\bfb$, so the model is fully described by $P(\bfA,\bfe,\bfb)=P(\bfA|\bfe,\bfb)P(\bfe|\bfb)P(\bfb)$, where $P(\bfA|\bfe,\bfb)$ is the model likelihood of NDC, $P(\bfe|\bfb)$ is the prior on edge counts and $P(\bfb)$ is the prior on the partition. For DC, we need to consider the additional prior on the degree sequence $\bfk$, and we thus have $P(\bfA,\bfe,\bfk,\bfb)=P(\bfA|\bfe,\bfk,\bfb)P(\bfk|\bfe,\bfb)P(\bfe|\bfb)P(\bfb)$ \cite{peixoto_nonparametric_2017}, where $P(\bfk|\bfe,\bfb)$ is the probability of the degree sequence. For PP, the prior on the edge counts serves as a constraint on the network to favour assortative structure. The full joint distribution can be written as $P(\bfA,\bfe,\bfk,\bfb)=P(\bfA|\bfe,\bfk,\bfb)P(\bfk|\bfe,\bfb)P(\bfe|\ein{},\eout{},\bfb)P(\ein{},\eout{}|E,\bfb)P(E)P(\bfb)$, where $E$ is the total number of edges in the network, $\ein{}$ and $\eout{}$ are the number of edges within and between blocks respectively, and where we use the model likelihood of DC \cite{zhang_statistical_2020}. 

We start with the elements of the joint distribution of NDC,
\begin{eqnarray}
    P(\bfA,\bfe,\bfb)=P(\bfA|\bfe,\bfb)P(\bfe|\bfb)P(\bfb)
\end{eqnarray}

The model likelihood of NDC is given by 
\begin{equation}
    P(\bfA|\bfe,\bfb) = \frac{\prod_{r<s}e_{rs}!\prod_re_{rr}!!}{\prod_rn_r^{e_r}\prod_{i<j}A_{ij}!\prod_iA_{ii}!!}
\end{equation}
The prior for the block matrix $e_{rs}$ is a uniform distribution over the total possible number of symmetric block matrices given $B$, with the constraint that the sum of all elements must equal $2E$:
\begin{equation}
    P(\bfe|\bfb) = \multiset{B(B+1)/2}{E}^{-1}
\end{equation}
The prior on the partition is defined as 
\begin{equation}
    \begin{aligned}
P(\bfb) &= P(\bfb|\bfn)P(\bfn|B)P(B) \\
     &= \frac{\sum_rn_r!}{N!} {N - 1 \choose B - 1}^{-1} \frac{1}{N}
\end{aligned}
\end{equation}
Here, $P(\bfb)$ and $P(\bfn|B)$ are hyperpriors on the number of blocks $B$ and on the block sizes $n_r$ respectively, to be as parsimonious as possible about these parameters.

In the case of DC, the model likelihood includes terms for the degree sequence $\bfk$, so that:
\begin{equation}
    P(\bfA|\bfe,\bfk,\bfb) = \frac{\prod_{r<s}e_{rs}!\prod_re_{rr}!!\prod_ik_i!}{\prod_re_r!\prod_{i<j}A_{ij}!\prod_iA_{ii}!!}
\end{equation}
Additionally, the DC case also includes a prior on the degree sequence $k$, namely 
\begin{equation}
    P(\bfk|\bfe,\bfb) = \prod_r\frac{\prod_k\eta_k^r!}{n_r!} \prod_r q(e_r, n_r)^{-1}
\end{equation}
where $\eta_k$ denotes the number of degree-$k$ nodes in group $r$ and $q(x,y)$ is the number of times an integer $x$ can be partitioned into a maximum of $y$ parts \cite{peixoto_nonparametric_2017}.

For PP, the prior on the block matrix needs to be defined differently, to encode the constraint that is put on the structural arrangement \cite{young_universality_2018}. In fact, Zhang and Peixoto \cite{zhang_statistical_2020} proposed two different versions of this probability: one which assumes that \emph{uniform} expected number of edges within each community and one that allows the number of expected edges to vary across communities (\emph{non-uniform}). Here, we give the formulation for both versions, since in our analysis, we use the uniform version in the case of synthetic networks (since they are generated with equal size blocks) and the non-uniform version in the analysis of the metablox vector on real networks. The uniform version of the prior on the edge counts in PP is described by
\begin{equation}
P(\bfe|\ein{},\eout{},\bfb)P(\ein{},\eout{}|E,\bfb)
\end{equation}
with 
\begin{equation}
    P(\bfe|\ein{},\eout{},\bfb) = \frac{\ein{}!\eout{}!}{B^{\ein{}}\prod_r(e_{rr}/2)! {B \choose 2}^{\eout{}}\prod_{r<s}e_{rs}!}.
\end{equation}
This is equivalent to the product of two uniform multinomial distributions: one for the elements of the block matrix that correspond to the within-block edge counts and one for those that correspond to the between-block edge counts, given $\ein{}$ and $\eout{}$. The second part is then the hyperprior on $\ein{}$ and $\eout{}$:
\begin{equation}
    P(\ein{},\eout{}|E,\bfb) = \left(\frac{1}{E+1}\right)^{1-\delta_{B,1}}
\end{equation}
For the non-uniform version, the prior is also made up of two probabilities
\begin{equation}
P(\bfe|\{e_{rr}\},\eout{},\bfb)P(\{e_{rr}\}, \eout{}|\bfb,E)
\end{equation}
where
\begin{equation}
    P(\bfe|\{e_{rr}\},\eout{},\bfb)=\frac{\eout{}!}{{B \choose 2}^{\eout{}}\prod_{r<s}e_{rs}!}
\end{equation}
corresponds to a uniform multinomial distribution for the off-diagonal elements of the block matrix given $\eout{}$. The second component is made out of a uniform distribution over all possible values $\ein{}$ from $E$, and a uniform distribution over all ways of choosing the set of diagonal block matrix values $\{e_{rr}\}$, given $\ein{}$:
\begin{equation}
\begin{aligned}
    P(\{e_{rr}\}, \eout{}|\bfb,E) &= P(\{e_{rr}\}|\ein{},\bfb)P(\ein{}|E,\bfb) \\
    &= {B + \ein{} -1 \choose \ein{}}^{-1} \left(\frac{1}{E+1}\right)^{1-\delta_{B,1}}
\end{aligned}
\end{equation}

\subsection*{\label{sec:sm_edgecompression}The role of network compressibility}
In the main text, we discussed the limitations of our measure and the way in which they can be mitigated by using the compression of the network under the \emph{optimal} per network edge, $c^{m}=\sopt / E$, as a second dimension.

Recall that we considered two limitations: one related to identifying the best SBM variant as part of a comparison of different dimensions of $\gamma$ and another regarding the conflation of block structure signals and metadata correlation. In Figures \textbf{4} and \textbf{5}, we used edge compression to highlight the particular SBM variant whose optimal partition provided the best partition of the network and therefore illustrated how this second partition can be used to address the first limitation. Including this additional information enabled a distinction between the cases in which the metadata is closest -- in terms of how well it compresses the network -- to \emph{the} optimal partition among the variants included in the analysis, or merely to the particular variant being measured. In other words, when the strongest relevance is measured under DC and DC is also the model whose optimal partition provides the best fit overall, we know that the metadata partition is strongly relevant not just under the particular model but in comparison to the most optimal partition we have for this network.

\begin{figure}
    \centering
    \includegraphics[width=0.9\linewidth]{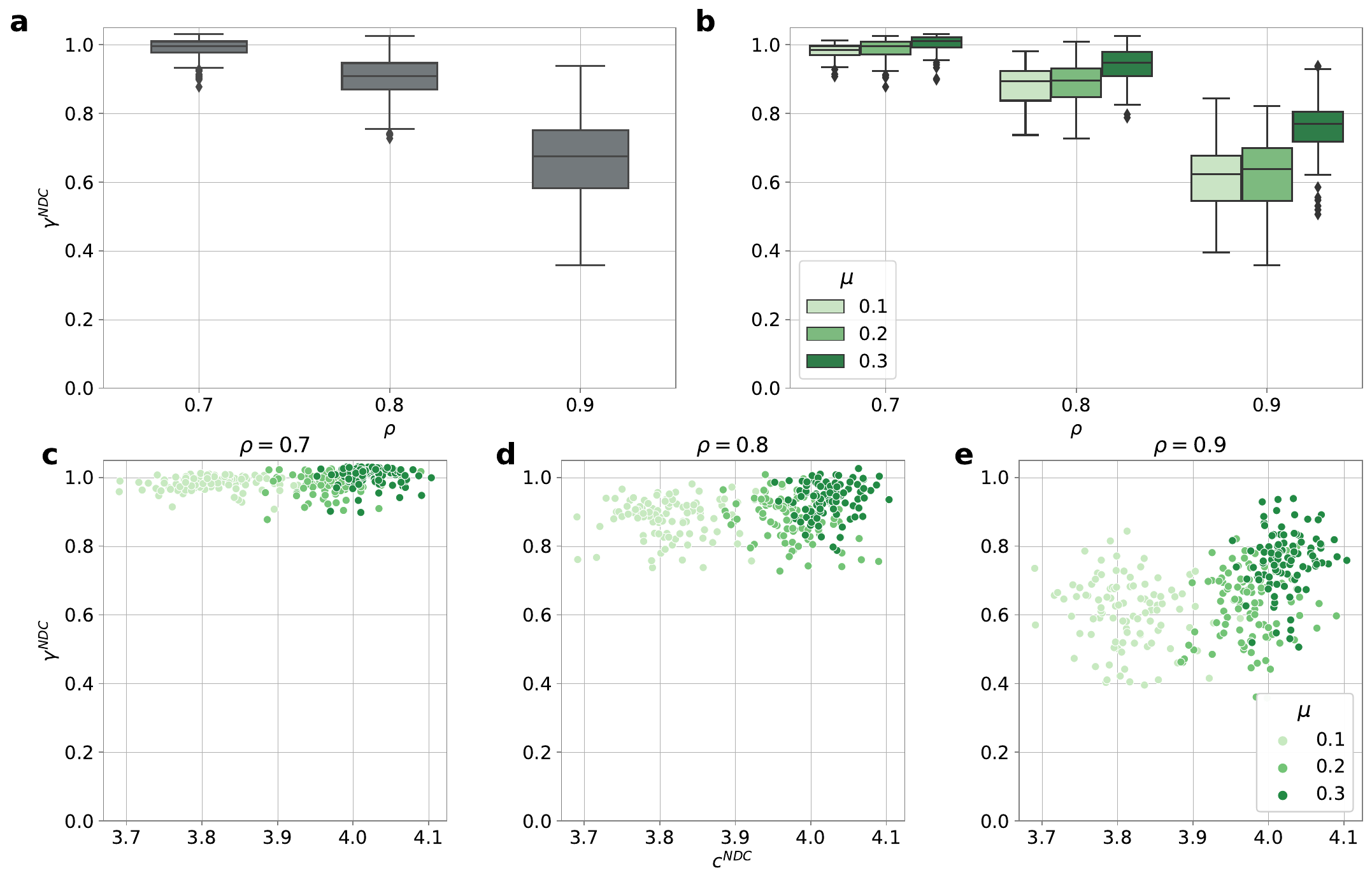}
    \caption[Using edge compression to disentangle block structure signal strength and metadata correlation.]%
    {Using edge compression to disentangle block structure signal strength and metadata correlation. \par \small 
    Distributions of non-degree-corrected metablox values for a total of $300$ networks, each with three sets of metadata. For box plots, the center line refers to the median, box limits to upper and lower quartiles and whiskers to 1.5 times the interquartile range; remaining points are outliers.
    \textbf{a} Box plots of the metablox values for three values of correlation $\rho$. 
    \textbf{b} Same box plots, disaggregated by block structure signal strength $\mu$.
    \textbf{c-e} Metablox values for the same data, as a function of edge compression $c$ under the non-degree-corrected variant, for correlation $\rho=0.7$ (\textbf{c}), $\rho=0.8$ (\textbf{d}), $\rho=0.9$ (\textbf{e}).}
    \label{fig:compressibility}
\end{figure}

Here, we provide a concrete example for the second limitation: on the ability of the second dimension to disentangle the signal of the block structure and the metadata correlation. For the purpose of this demonstration, we generate a total of $300$ synthetic networks with community structure, for each of which we fix size $N=200$, expected degree $k=10$ and number of blocks $B$. We create the networks in this collection such that we end up with three sets of $100$ networks with differently `strong' community structure. The stronger the signal of the community structure, the more pronounced the difference between the within- and between-block edge probability. Specifically, we generate the networks using an SBM with block matrix $\boldsymbol{\theta}_{\text{BC}}=2E\big(\begin{smallmatrix} 1 - \mu & \mu \\ \mu & 1 - \mu \end{smallmatrix}\big)$, with $\mu=0.1$ for the strongest signal, $\mu=0.2$ for a medium signal and $\mu=0.3$ for weak community structure. For each network, we generate three sets of categorical node metadata, correlating with the planted block structure with correlation $\rho=0.7; 0.8; 0.9$. In Figure \textbf{6}, we visualise an analysis of the NDC element of metablox for these networks; we refrain from including it for other dimensions as those yielded very similar results. In the panel on the left-hand side of Figure \textbf{6a}, we see that increasing metadata correlation leads to lower values of $\gamma$. For Figure \textbf{6b}, we disaggregated the data by signal strength. We observe that for each correlation value $\rho$, the $\gamma$ values for the networks with strong and medium community structure are essentially indistinguishable, while the networks with weak community structure have a larger value of $\gamma$ for the same $\rho$. While $\gamma$ on the whole does a good job at differentiating between the correlation strength of the different network-metadata pairs, the differences we observe upon separating by community strength call for further analysis. Especially in light of working with real networks, whose generative process is unknown to us, these insights demand a tool to disentangle the different impacting factors on the metadata relevance. In Figures \textbf{6c-e}, we demonstrate the way in which the edge compression mitigates this limitation: As expected, the stronger the community structure, the `better' (\hbox{i.e.} smaller $c$) the compression of the optimal partition per network edge.

\begin{figure*}
\centering
\includegraphics[width=0.9\textwidth]{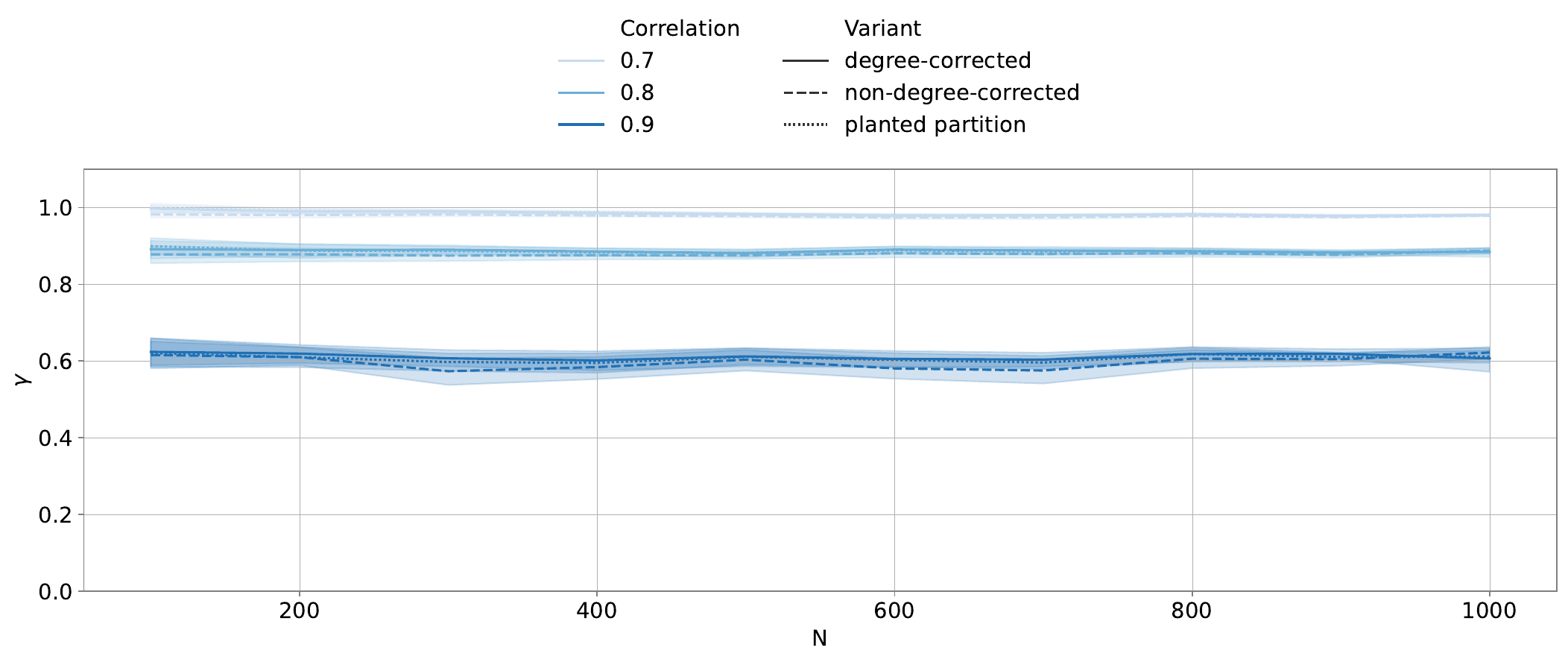}
    \caption[Metablox values for networks of varying size.]%
    {Metablox values for networks of varying size. \par \small 
    Metablox values for networks of varying size $N$, for three different correlation values $\rho$ and three variants (degree-corrected, non-degree-corrected, and planted partition). The lines show the mean metablox values across $50$ networks, the shaded areas show $95\%$ confidence intervals.}
\label{fig:robustness_N}
\end{figure*}

\subsection*{\label{sec:sm_robustness}Robustness analysis}
To understand the suitability of metablox to be used as part of comparative studies, we analyse the sensitivity of the measure to networks of different sizes and topologies. We first discuss the parameters to which we do and do not expect the measure to be robust. For an inter-network comparison, it is important that our measure is robust with respect to varying network size $N$, which is not a trivial question since the description length of a network increases with growing $N$. For structural differences between two networks that are related to the compressibility under an SBM, such as the number of blocks, the edge density, or the signal strength of the block structure, the question becomes more nuanced. In fact, we have already discussed in `Limitations and network compressibility' that -- due to our definition of metadata-block structure relevance -- we \emph{expect} the measure to be impacted by the compressibility of a network: metadata is more relevant to a network with stronger block structure signal. We therefore expect metablox to yield stronger relevance for network-metadata pairs for which the SBM variant provides better compression (\hbox{e.g.} for increasing numbers of blocks.

\begin{figure}
    \centering
    \includegraphics[width=0.9\linewidth]{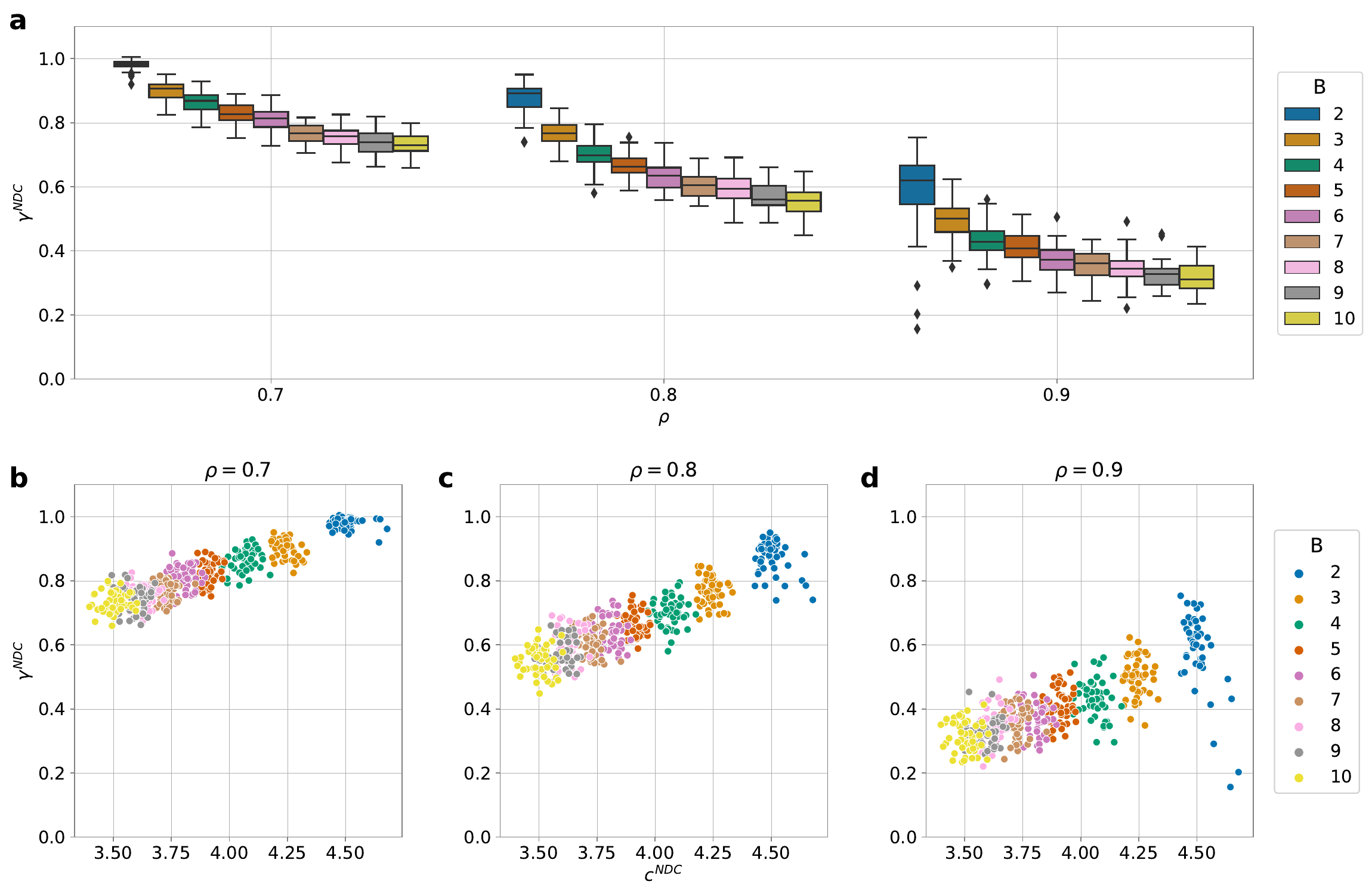}
        \caption[Metablox values for varying number of blocks.]%
        {Metablox values for varying number of blocks. \par \small 
        Distributions of non-degree-corrected metablox values for networks of size $N=400$ with varying number of blocks $B$. $50$ networks were created for each $B$ and metadata was created with varying levels fo correlation $\rho$ with the planted block structure. For box plots, the center line refers to the median, box limits to upper and lower quartiles and whiskers to 1.5 times the interquartile range; remaining points are outliers.
        \textbf{a} Box plots of the metablox values for three values of correlation $\rho$, disaggregated by number of blocks $B$.
        \textbf{b-d} The same metablox values as a function of edge compression $c$ under the non-degree-corrected variant, for correlation $\rho=0.7$ (\textbf{b}), $\rho=0.8$ (\textbf{c}), $\rho=0.9$ (\textbf{d}).}
    \label{fig:robustness_B}
\end{figure}

To test these assumptions, we generate a number of synthetic networks. Specifically, we increase $N$ from $N=100$ to $N=1000$ at steps of $100$, and generate $50$ networks for each $N$, all with expected degree $k=10$ and with planted bicommunity structure ($B=2$). For the within- and between block edge counts, we use the same block matrix $\boldsymbol{\theta}_{\text{BC}}$ as above, with $\mu=0.1$. In Figure \textbf{7}, we plot the mean description lengths and metablox (plus $95\%$ confidence intervals) for each value of $N$ and three SBM variants, for correlation values of $\rho=0.7, 0.8, 0.9$. We clearly see the intended normalising effect, as $\gamma$ remains stable for growing $N$ for all three variants.

To test the measure's behaviour for varying numbers of blocks, we fix the network size at $N=400$ and leave all other parameters as above, increasing $B$ from $2$ to $10$, again generating $50$ networks at each step. In Figure \textbf{8a} we show the results for $\gndc$ (the results for the other two variants are similar). We observe that, for each value of $\rho$, $\gamma$ decreases with the number of blocks, confirming our expectation of the measure capturing stronger metadata relevance for more compressible networks. In Figures \textbf{8b-d}, we show how the edge compression of the network can be used as a second dimension, if a distinction between metadata correlation and block structure signal strength is required. The three figures show the edge compression $c$ on the x-axis and $\gamma$ on the y-axis. A lower value of $c$ implies better compression, and we can see that the second dimension separates the compressibility from the metadata correlation by placing the most compressible networks low on the x-axis and low on the y-axis. In a network comparison, in which disentangling these two factors is important, researchers can plot the two dimensions in this way to draw the correct conclusions.

\subsection*{\label{sec:landscape}Heterogeneous partition landscape}
For the synthetic network used to illlustrate the measure in the Results section, we demonstrate here that this network does, in fact, exhibit a heterogeneous partition landscape -- as designed by the SCBM \cite{mangold_generative_2023}. Recall that we created a network with $N=100$ nodes, $B=2$ blocks and expected degree $k=10$, generated in a way such that upon fitting an SBM to the network and sampling from the posterior distribution, one recovers at least two `clusters' of partitions -- a bicommunity partition (BC) and a second one dividing the network into a core and a periphery (CP). We choose $\boldsymbol{\theta}_{\text{BC}}=2E\big(\begin{smallmatrix} 1 - \mu & \mu \\ \mu & 1 - \mu \end{smallmatrix}\big)$ and $\boldsymbol{\theta}_{\text{CP}}=2E \big(\begin{smallmatrix} 1 - \lambda & \frac{1}{2} \\ \frac{1}{2} & \lambda \end{smallmatrix}\big)$ as block matrices, with $\mu=0.25$ and $\lambda=0.05$ and where $E$ denotes the total number of edges. The choice of $\mu$ and $\lambda$ is motivated by the findings in Ref \cite{mangold_generative_2023}, where it was demonstrated that the partition landscape inferred by a degree-corrected SBM on a network with two partitions planted with these parameters does, in fact, uncover both the planted partitions. The probability of the existence of edges in this network depends solely on the block membership of nodes and on the edge probability given by these two matrices. 

\begin{figure*}
\centering
\includegraphics[width=0.9\linewidth]{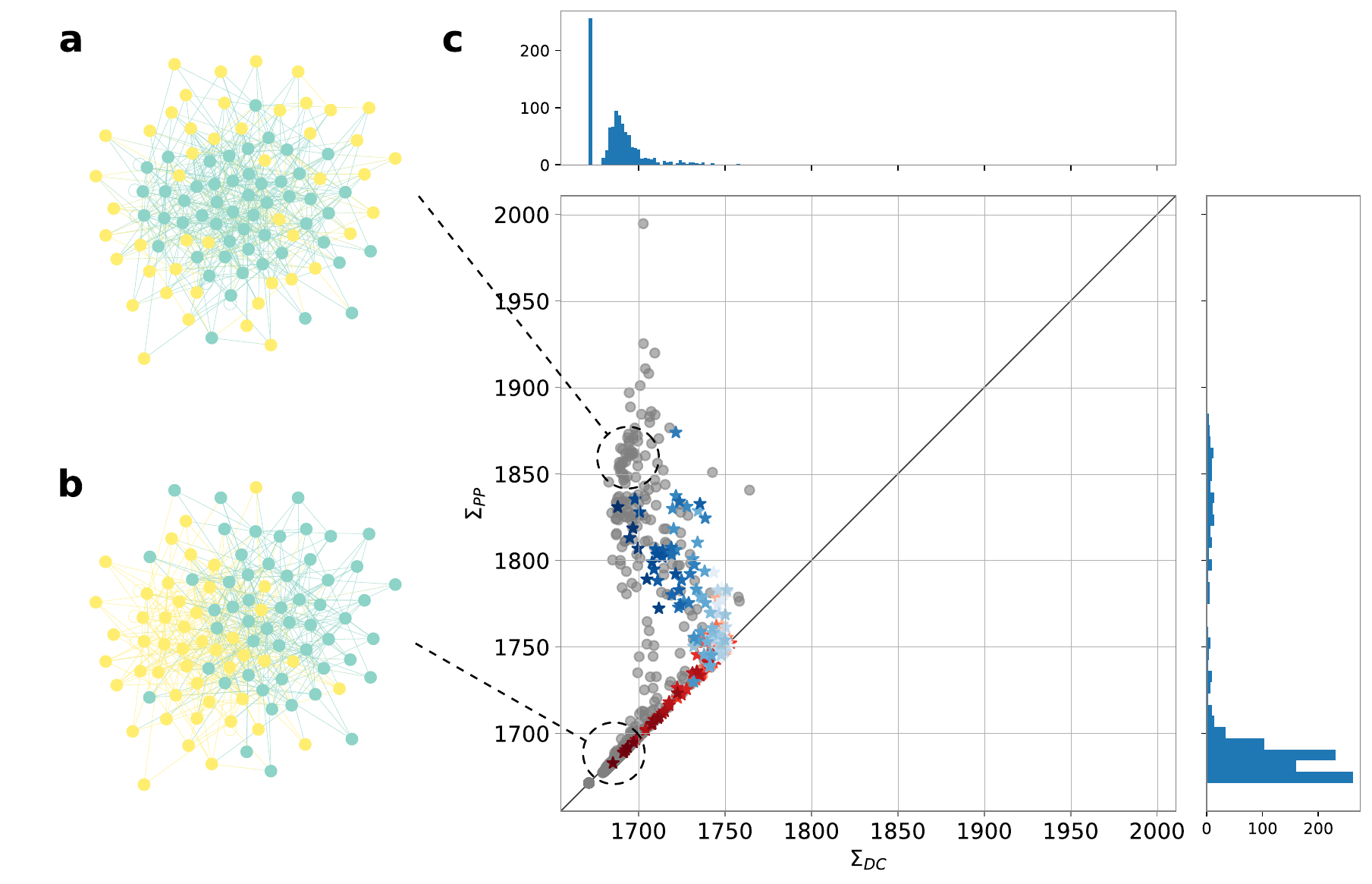}
    \caption[Heterogeneity of the partition landscape.]%
    {Heterogeneity of the partition landscape. \par \small 
    \textbf{a} Visualisation of the synthetic network described in the main text, with nodes painted according to a partition that resembles the planted core-periphery structure. 
    \textbf{b} The same network, with nodes painted according to a partition that resembles the planted bicommunity structure. 
    \textbf{c} Description lengths under the degree-corrected and planted partition Stochastic block model of the inferred partitions (grey dots), sampled by fitting a degree-corrected Stochastic block model, and description lengths of the same network under a degree-corrected and planted partition variant, with the partition parameter set equal to different sets of metadata partitions. Red (blue) stars represent description lengths under the bicommunity-like (core-periphery-like) metadata. The darker the colour of the stars representing the metadata partitions, the higher the correlation of the metadata labels with the respective planted structure. The insets show histograms of the description lengths of the inferred partitions.}
\label{fig:schem_combined}
\end{figure*}

For the purpose of demonstrating that this network's partition landscape does have multiple competing explanations, we fit an SBM and sample from the posterior distribution by using the methods from the graph-tool library \cite{peixoto_graph-tool_2014}. In particular, we use the degree-corrected SBM variant (DC), which is meant to account for heterogeneous degree distributions within blocks \cite{karrer_robustness_2008}. The description lengths $\Sigma^{\text{DC}}$ of the network according to these partitions are plotted in grey dots on the x-axis in Figure \textbf{9c}. We also calculate the description lengths $\Sigma^{\text{PP}}$ of the network according to each partition under the planted partition SBM (PP), the variant of the model that assumes assortativity \cite{zhang_statistical_2020}; these are shown on the y-axis of the same figure. Note that we have $\Sigma^{\text{PP}} \geq \Sigma^{\text{DC}}$ for all partitions, since partitions were found by DC and corresponding description lengths were then calculated for the same partitions under PP. If we take a closer look at the partitions that we sampled, we find groups of similar partitions, here marked by the dashed-line circles, that look like the two partitions we planted. We show representative partitions for each partition cluster in the two network visualisations in Figures \textbf{9a} and \textbf{9b}, in which nodes are painted according to block assignments. We clearly see a strong similarity between these inferred partitions and each of the two planted partitions in Figure \textbf{1} and we therefore conclude that we have successfully created a network with the desired diverse partition landscape. We also observe that for those inferred partitions that are similar to the planted bicommunity partition, the PP offers a similarly good encoding of the network as the DC, since $\Sigma^{\text{PP}} \approx \Sigma^{\text{DC}}$ for those partitions. In contrast, the description lengths of the network are considerably higher under the same partitions but according to PP. This illustrates that calculating the description length of a network under different models can help us probe its partition landscape. 

In line with the objective of our measure, we use this example network  to support our hypothesis that using description length is a suitable tool to understand the way in which metadata is relevant for different parts of the partition landscape. For this purpose, we calculate $\Sigma_d^{\text{DC}}$ and $\Sigma_d^{\text{PP}}$ for each of the $202$ sets of metadata that we generated for this network, as introduced in the Results section and plot the resulting description length values alongside the grey dots (\hbox{i.e.} alongside the description length values of the \emph{inferred} partitions) on Figure \textbf{9c}. The red stars represent the metadata partitions that are correlated with the planted bicommunity structure, the blue stars represent those that are similar to the core-periphery structure, with darker colours depicting higher values of $\rho$. As expected, we observe that under the more general of the two models, the partitions with the highest values of $\rho$ have the lowest description length: the stronger the correlation of the metadata with the planted structure, the lower the description length, since the metadata partition is similar to the two ground truth partitions that were responsible for generating the network. The description lengths under the PP on the y-axis illustrate that additional to measuring the extent to which metadata is related to structure, we can also probe the type of structural arrangement: the bicommunity-like metadata are encoded as well under PP as under DC, indicating that assortativity was a prominent feature of the network generation process \cite{zhang_statistical_2020}. The core-periphery-like metadata, however, yield much higher description lengths under PP compared to DC, suggesting that -- as we know is true -- if we were to assume that this metadata was responsible for generating this network, assortativity was much less likely the prominent structure compared to some other more general structure. 

\newpage

\section*{Code availability} 
A Python library to use the \emph{metablox} measure can be downloaded and installed from \url{https://github.com/lenafm/metablox}. The scripts to generate the results in the paper are available in \url{https://github.com/lenafm/quantifying-metadata-relevance}.

\section*{Data Availability}
The data on law firm networks \cite{lazega_collegial_2001} is available in ``The Netzschleuder network catalogue and repository'' (doi:10.5281/zenodo.7839981) \cite{peixoto_netzschleuder_2020}. The impact investing data is available in the NAKALA repository (doi:10.34847/nkl.dbd8q853) \cite{data_impact_investing}. The scripts to generate all other data are available in \url{https://github.com/lenafm/quantifying-metadata-relevance}.

\section*{Acknowledgements}
We are thankful to Manuel Tonneau for his contribution to the Impact Investing data collection endeavor, as well as Ève Chiapello for the initial impulse in investigating this field on Twitter. This work was partially supported by the ``Socsemics" Consolidator grant from the European Research Council (ERC) under the European Union’s Horizon 2020 research and innovation program (grant agreement No. 772743).

\section*{Author contributions statement}
L.M. and C.R. designed the analysis. L.M. wrote the code, performed the analysis, and prepared the figures. L.M. and C.R. wrote and approved the manuscript.

\section*{Ethics declarations}
\textbf{Competing interests:} The authors declare that they have no competing interests.

\newpage


\begin{sloppypar}
\printbibliography[title={References}]
\end{sloppypar}

\end{document}